\PassOptionsToPackage{twoside=false}{geometry}
\documentclass[manuscript,screen]{acmart}
\AtBeginDocument{%
  }

\setcopyright{acmlicensed}
\copyrightyear{2018}
\acmYear{2018}
\acmDOI{XXXXXXX.XXXXXXX}

\acmJournal{JACM}
\acmVolume{37}
\acmNumber{4}
\acmArticle{111}
\acmMonth{8}

\usepackage{algorithmic}
\usepackage{graphicx}
\usepackage{textcomp}
\usepackage{xcolor}
\usepackage{multirow}
\usepackage{tabularx}
\usepackage{url}
\usepackage{hyperref}
\usepackage{makecell}
\usepackage{graphicx}
\usepackage{subcaption}
\usepackage{pifont}
\usepackage{tcolorbox}
\usepackage{xcolor}
\usepackage[para,online,flushleft]{threeparttable}
\newcommand{\cmark}{\ding{51}}%
\newcommand{\xmark}{\ding{55}}%

\usepackage{listings}
\begin{document}

\title{Dissecting the NVIDIA Hopper Architecture through Microbenchmarking and Multiple Level Analysis}\thanks{}


\author{Weile Luo}
\affiliation{%
  \institution{The Hong Kong University of Science and Technology (Guangzhou)}
  \city{Guangzhou}
  \country{China}}
\email{w.luo@connect.hkust-gz.edu.cn}

\author{Ruibo Fan}
\affiliation{%
  \institution{The Hong Kong University of Science and Technology (Guangzhou)}
  \city{Guangzhou}
  \country{China}}
\email{rfan404@connect.hkust-gz.edu.cn}

\author{Zeyu Li}
\affiliation{%
  \institution{The Hong Kong University of Science and Technology (Guangzhou)}
  \city{Guangzhou}
  \country{China}}
\email{zli755@connect.hkust-gz.edu.cn}

\author{Dayou Du}
\affiliation{%
  \institution{The Hong Kong University of Science and Technology (Guangzhou)}
  \city{Guangzhou}
  \country{China}}
\email{dda487@connect.hkust-gz.edu.cn}

\author{Hongyuan Liu}
\affiliation{%
  \institution{The Hong Kong University of Science and Technology (Guangzhou)}
  \city{Guangzhou}
  \country{China}}
\email{hongyuanliu@hkust-gz.edu.cn}

\author{Qiang Wang}
\affiliation{%
  \institution{Harbin Institute of Technology, Shenzhen}
  \city{Shenzhen}
  \country{China}}
\email{qiang.wang@hit.edu.cn}

\author{Xiaowen Chu}
\affiliation{%
  \institution{The Hong Kong University of Science and Technology (Guangzhou)}
  \city{Guangzhou}
  \country{China}}
\email{xwchu@hkust-gz.edu.cn}

\renewcommand{\shortauthors}{Weile et al.}

\thanks{Qiang Wang and Xiaowen Chu are the corresponding authors.}


\begin{abstract}
This study presents a comprehensive multi-level analysis of the NVIDIA Hopper GPU architecture, focusing on its performance characteristics and novel features. We benchmark Hopper's memory subsystem, highlighting improvements in the L2 partitioned cache and global memory access compared to Ampere and Ada Lovelace. The evaluation of Hopper's fourth-generation tensor cores reveals the benefits of FP8 precision and asynchronous \texttt{wgmma} instructions for matrix operations. Additionally, we investigate the performance of DPX instructions for dynamic programming, distributed shared memory (DSM) for inter-SM communication, and the Tensor Memory Accelerator (TMA) for asynchronous data movement. Through multi-level evaluation, we discover that the Hopper architecture demonstrates significant acceleration potential in real-world applications. For instance, the asynchronous programming model supported by TMA achieves a 1.5× speedup in matrix multiplication, FP8 delivers nearly double the performance of FP16, and DPX instructions accelerate a computational biology algorithm by at least 4.75×. Our findings provide actionable insights for optimizing compute-intensive workloads, from AI training to bioinformatics, on Hopper GPUs.
\end{abstract}



\received{20 February 2007}
\received[revised]{12 March 2009}
\received[accepted]{5 June 2009}

\def\thefootnote{}\footnotetext{This work is based on and extends "Benchmarking and Dissecting the Nvidia Hopper GPU Architecture" published in IPDPS 2024. This article expands previous work by adding new tests on L2 partitioned cache, energy consumption of tensor core instructions, comprehensive distributed shared memory evaluations, application-level DPX instruction tests, and benchmarking TMA performance under asynchronous programming on Hopper.  The artifact is available on \url{https://github.com/HPMLL/NVIDIA-Hopper-Benchmark}.}

\maketitle

\section{Introduction}
The increasing prevalence of GPUs has significantly accelerated a diverse range of computational workloads, from scientific computing to artificial intelligence, particularly with the rise of large language models (LLMs) such as GPT-3, with over 150 billion parameters \cite{floridi2020gpt}. Modern GPU architectures such as Ampere, Ada, and Hopper, incorporate specialized hardware like tensor cores and high-bandwidth memory systems, purpose-built for AI applications. As a result, GPUs have become indispensable components in high-performance computing clusters.

NVIDIA maintains a two-year release cycle for new GPU architectures, ensuring the continuous integration of novel architectural features. However, the scarcity of detailed micro-architectural specifications hinders precise performance analysis, necessitating deeper investigations to understand the impact of these advancements on application performance. Such comprehensive architectural understanding is fundamental for several critical applications: accurate performance and power modeling for energy-efficient computing\cite{isca2010_hong,taco2021_portable,hpca2019_hybrid,tpds2020_gpudvfs,mei2017survey,arafa2020verified,van2022isolating,arafa2019low}, the development of robust GPU simulators that enable cost-effective research and development \cite{accelsim_isca2020,bakhoda2009analyzing,gpuwattch_isca2013}, and the effective optimization of applications across diverse domains from AI training to scientific computing \cite{bakhoda2009analyzing,ho2017exploiting,yan_optimize_ppopp2020}. Without detailed microarchitectural insights, developers cannot fully exploit the potential of modern GPUs, leading to suboptimal performance and energy efficiency.

The evolution of tensor cores (TCs) exemplifies this trend. Introduced in the Volta architecture to accelerate deep neural networks using FP16/FP32 operations, TCs have been progressively enhanced. Ampere and Hopper continue to increase support for different precisions, providing more space for research on LLM. 
Thus, the reliance on assembly analysis and microbenchmarks from older architectures (Ampere and Turing) highlights the need for Hopper-specific TC research.


Beyond enhanced TCs, the Hopper architecture introduces several key innovations. Dynamic Programming X (DPX) instructions accelerate dynamic programming algorithms, which often rely on comparisons (min/max operations) between previously computed results. Distributed shared memory (DSM) enables direct communication between Streaming Multiprocessors (SMs), facilitating loads, stores, and atomic operations across their respective shared memory blocks. Finally, the Tensor Memory Accelerator (TMA) provides an asynchronous data movement mechanism between hardware components. However, the implementation details and performance characteristics of the TMA remain largely unexplored in the existing literature. A comprehensive understanding of these aspects is crucial for developers seeking to optimize compute-intensive workloads, enable accurate performance modeling, and design efficient algorithms that can fully exploit the potential of modern GPUs. Further research is needed to fully understand the interplay of these features and their combined effect on diverse workloads.

This article extends our previous conference paper \cite{10579250}, which explored the basic performance of the memory subsystem, a multi-level comparison of fourth-generation tensor cores with previous architectures, and a simple analysis of DPX and Distributed Shared Memory. In this work, we expand both the depth and breadth of our research. 
In terms of depth, we further analyze the memory subsystem. After various comparisons, we find that some traditional benchmarks are not suitable for Ampere and Hopper. We conduct new tests on the L2 partitioned cache for Ampere and Hopper, overturning previous data and conclusions. Additionally, we include energy consumption tests for \texttt{wgmma} in the tensor core, revealing the power impact of new instructions. We also perform a comprehensive evaluation of distributed shared memory (DSM), assessing DSM interface costs, expanding cluster numbers in latency tests, and testing throughput with different access patterns and thread block scheduling policies. Furthermore, we incorporate application-level testing to further demonstrate the advantages and limitations of DPX instructions.
In terms of breadth, with the rise of asynchronous programming, we benchmark the TMA on Hopper to characterize its performance within asynchronous programming paradigms.

In this study, we conduct a comprehensive multi-level benchmarking of the latest GPU architectures (Ampere, Ada, and Hopper). Additionally, we test Hopper's new features (DSM, TMA, and DPX). To our knowledge, this is the most comprehensive and in-depth benchmarking of the Hopper architecture. Our research presents a pioneering analysis of the new programming interfaces specific to the Hopper architecture, offering a unique horizontal performance comparison among these cutting-edge GPU architectures. Many of our findings are novel and being published for the first time, providing valuable insights.
We highlight the contributions of our work as follows: 

    \vspace{0.2em}
    $\bullet$ We systematically benchmark the latency and throughput of Hopper GPUs, including the memory subsystem, tensor cores, distributed shared memory (DSM), tensor memory accelerator (TMA), and dynamic programming instructions (DPX). 
    We compare the latency and throughput of these Hopper-specific units with their predecessors in Ampere and Ada Lovelace architectures, highlighting the performance improvements and trade-offs introduced in Hopper.

    
    $\bullet$ We perform a \emph{multi-level analysis} of Hopper's critical components, benchmarking them at the instruction, library, and application levels.  
    For the tensor core benchmark, we include tests at the library level of the transformer engine and real LLM generation applications. In DSM, we consider different access patterns in high-performance computing to provide valuable references for real applications. In both DSM and DPX, we employ real-world applications for evaluation to explore the practical potential of these new features.

    $\bullet$ Our research explores the unique features of the Hopper architecture comprehensively, including the L2 partitioned cache, \texttt{wgmma} instructions, DPX, TMA, and DSM. 
    These innovations have the potential to significantly enhance GPU programming methodologies. 
    By evaluating their performance, our study contributes to performance modeling and algorithm design in dynamic programming and scientific computing, potentially unlocking the full performance potential of GPUs and driving advancements in the field.

The remainder of this paper is organized as follows. Section~\ref{sec:works} reviews related work. Section \ref{sec:hopper} briefly introduces the Hopper architecture and its new features. In Section~\ref{sec:mem}, we delve into the latency and throughput of the GPU memory subsystem. Section \ref{sec:tma} examines the latency and throughput of the tensor memory accelerator. Section~\ref{sec:tc} evaluates the tensor core at the instruction level, the transformer engine at the library level, and LLMs at the application level. Sections~\ref{sec:dsm} and \ref{sec:dpx} respectively benchmark distributed shared memory and DPX instructions. 
Finally, Section 9 summarizes the findings.

\section{Related Work}
\label{sec:works}

The rapid evolution of GPU architectures necessitates continuous investigation into their microarchitectural details, particularly at the instruction level.  This understanding is fundamental for accurate performance and power modeling \cite{isca2010_hong,taco2021_portable,hpca2019_hybrid,tpds2020_gpudvfs,mei2017survey,arafa2020verified,van2022isolating,arafa2019low}, the development of robust simulators \cite{accelsim_isca2020,bakhoda2009analyzing,gpuwattch_isca2013}, and the effective optimization of applications \cite{bakhoda2009analyzing,ho2017exploiting,yan_optimize_ppopp2020}.  Prior research has addressed these needs across different GPU generations, but the introduction of novel features in recent architectures, such as Hopper, demands a renewed focus.  

Early efforts to dissect GPU microarchitecture primarily focused on older generations like Fermi, Kepler, and Maxwell \cite{wong2010demystifying,mei2014benchmarking,mei_tpds_2017}.  These studies, along with investigations into performance and power characteristics \cite{jia2018dissecting,jia2019dissecting} and software-level scheduling frameworks\cite{8853389}, laid the groundwork for understanding fundamental GPU behavior and informed subsequent investigations into more recent architectures like Volta and Turing.  The emergence of tensor cores (TCs) in Volta, specialized hardware units designed to accelerate matrix operations crucial for AI workloads, marked a significant shift in GPU architecture and spurred a new wave of research.

The importance of fused matrix multiplication accumulation (MMA) in AI has driven extensive research into TCs.  Initial studies on Volta's TCs centered on the legacy \texttt{wmma} API and employed benchmarks using established libraries like CUBLAS and CUTLASS \cite{markidis_tensor_ipdpsw2018,benchmark_Martineau_eupar2019}.  Further work extended this analysis to Turing, incorporating preliminary assembly code analysis of \texttt{wmma} instructions \cite{jia2018dissecting,jia2019dissecting}.  However, these initial explorations lacked the comprehensive instruction-level microbenchmarks needed to fully characterize TC performance and numerical behavior.

Subsequent research delved deeper into TC analysis on Volta and Turing, utilizing assembly-level (SASS) benchmarking to uncover optimization opportunities for matrix multiplication, particularly in half-precision \cite{yan_ipdps_2020,yan_optimize_ppopp2020,raihan_ispass2019}.  The numerical intricacies of TCs, including rounding modes and subnormal number handling for various data types like TF32, BF16, and FP16, were also investigated \cite{fasi2021numerical}.  However, the limitations of the \texttt{wmma} API, particularly its inability to fully leverage new features like sparse matrix multiplication introduced in Ampere and later architectures, became increasingly apparent  \cite{markidis_tensor_ipdpsw2018, fasi2021numerical}.

The introduction of the \texttt{mma} API with Turing addressed some of these limitations and evolved to support sparse matrix multiplication (\texttt{mma.sp}) on Ampere and beyond.  Sun et al. \cite{sun_tpds_2023} conducted a thorough analysis of the \texttt{mma} API on Turing and Ampere, exploring TC performance, numerical behavior, and sparse matrix multiplication capabilities.  However, the arrival of Hopper introduced further complexity with the addition of \texttt{wgmma} and an expanded set of \texttt{mma} instructions supporting a broader range of precisions.  This necessitates new research to understand the performance characteristics and optimal utilization of these new instructions, a gap addressed by our work.

Beyond TC-specific research, previous work has also addressed energy efficiency in GPUs using static \cite{hong2009analytical, isca2010_hong, braun2020simple, fan2019predictable} and dynamic \cite{tpds2020_gpudvfs, guerreiro2018gpgpu, guerreiro2019modeling} analysis techniques, as well as hybrid approaches \cite{wang2024dso}. Application-level benchmarks have explored the impact of DVFS on deep learning workloads \cite{tang2019impact, wang2020benchmarking}. 

\begin{table*}[!h]
	\centering
	\caption{The Comparison of the studies of GPGPU Microbenchmark. 
Items in parentheses indicate expansions over \cite{10579250}.}
	\label{tab:mb_compare}
	\scriptsize{
            \setlength{\tabcolsep}{2pt}
		\begin{tabular}{|c|c|c|c|c|c|c|c|} \hline
			\textbf{Study} 
			& \textbf{Instruction-level} & \textbf{Memory subsystem} &\textbf{Energy  consumption} & \textbf{Tensor core} & \textbf{Application evaluation} & \textbf{Architectural features} & \textbf{Target architecture}\\ \hline	
			\cite{jia2018dissecting,markidis_tensor_ipdpsw2018,benchmark_Martineau_eupar2019}
			& \cmark  & \cmark & \xmark  & \cmark & \xmark & \xmark & Volta \\ \hline 
                \cite{jia2019dissecting,raihan_ispass2019,yan_ipdps_2020} 
                & \cmark  & \cmark & \cmark  & \cmark & \xmark & \xmark & Turing,Volta  \\ \hline
                \cite{fasi2021numerical,heart_ampere_2021,abdelkhalik_hpec_2022,sun_tpds_2023} 
                & \cmark  & \cmark & \xmark  & \cmark & \xmark & Asynchronous Copy & Ampere,Turing  \\ \hline
			\cite{10579250} & \cmark  & \cmark & \cmark  & \cmark & \cmark & DSM, DPX & Ampere, Ada, Hopper \\ \hline 
                This study & \cmark  & \cmark (L2 partitioned cache) & \cmark (\texttt{wgmma}) & \cmark & \cmark(DSM, TMA, DPX)  & DSM, TMA, DPX & Ampere, Ada, Hopper \\ \hline 
			
		\end{tabular}
    	}
\end{table*}

Table \ref{tab:mb_compare} summarizes the existing microbenchmark studies across different GPU architectures and compares with our work. In contrast to previous research, our study provides a holistic analysis of Hopper, including its novel TCs, DPX instructions, DSM, and TMA.  As Table \ref{tab:mb_compare} demonstrates, our work provides comprehensive benchmarks, offering valuable insights for developers seeking to optimize performance and energy efficiency on this latest architecture.
\section{Overview of Hopper Architecture}
\label{sec:hopper}
Hopper, as a next-generation GPU architecture, offers enhancements in two areas. One involves strengthening traditional GPU functions such as peak performance, cache, memory, and tensor cores. The other focuses on designing new features to address evolving real-world tasks, such as the TMA, DSM, and DPX. In this section, we introduce these new aspects of Hopper and benchmark them in subsequent sections.

\begin{figure*}[ht]
    \centering
    \includegraphics[width=\linewidth]{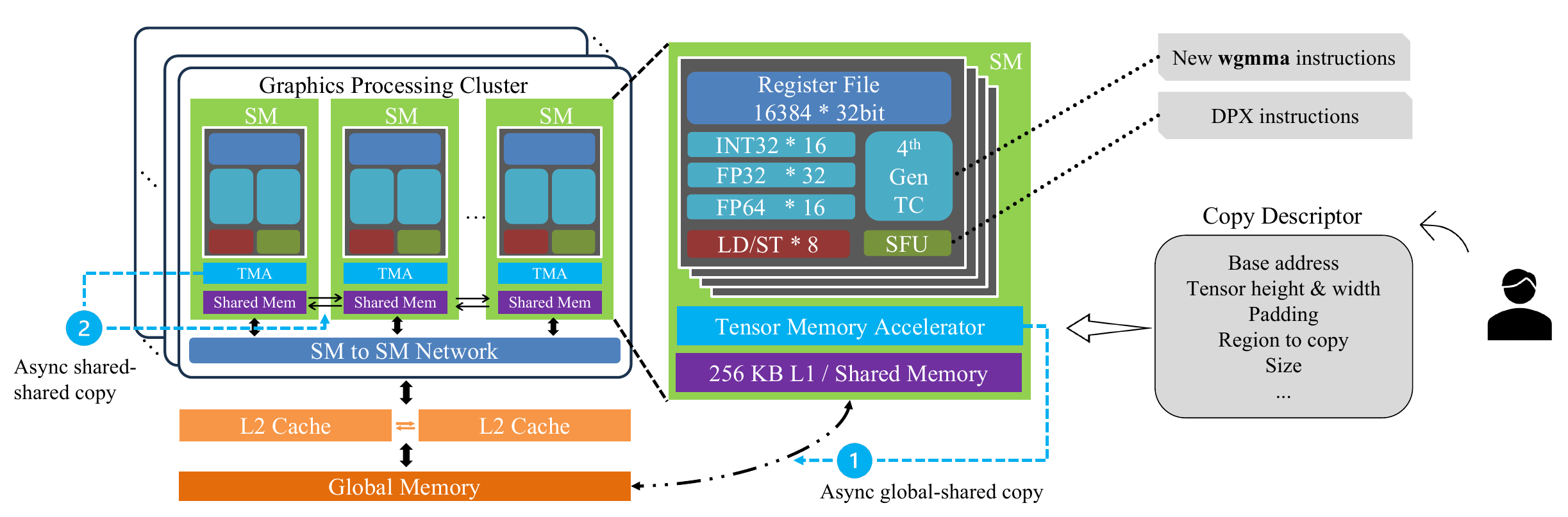} 
    \caption{Hopper architecture and new features}
    \label{fig:hopper}
    \vspace{-1.0 em}
\end{figure*}

The new Hopper architecture features several key enhancements, as illustrated in Fig. \ref{fig:hopper}:
\begin{itemize}
    \item Hopper features a \textbf{50MB partitioned L2 cache}, a 25\% increase compared to the 40MB in Ampere, retaining the partitioned design introduced in the previous generation. 
    \item \textbf{Fourth-generation tensor cores} offer improved computational performance, introducing new \texttt{wgmma} instructions on Hopper for asynchronous operations and FP8 format support, benefiting large-scale deep neural network models.
    \item \textbf{DPX instructions}, supported by hardware acceleration on Hopper, facilitate dynamic programming algorithms by enabling fast max/min calculations with fused add operations and ReLU (clamping to zero).
    \item A new \textbf{thread block cluster} level abstracts to the hardware Graphic Processing Cluster (GPC), allowing SM-to-SM communication via the SM-to-SM network within a GPC.
    \item New \textbf{asynchronous execution features} include the \textbf{(TMA)} and an asynchronous transaction barrier.
\end{itemize}

In this work, all the above new features will be benchmarked in multiple levels. We select the most representative GPUs of the Ampere, Ada Lovelace, and Hopper architectures, which are A100 PCIe, RTX~4090, and H800 PCIe respectively. Their basic hardware properties are shown in Table \ref{tab:hw_device}. 
In terms of software configurations, the RTX~4090 utilized driver version 530.30.02 and CUDA version 12.1, while the A100 and H800 employed driver version 560.35.03 and CUDA version 12.6.

\begin{table}[ht]
    \centering
    \caption{Comparison of the Properties of the Ampere, Ada Lovelace and Hopper Devices}
    \label{tab:hw_device}
    
    \begin{tabular}{l ccc}
        \toprule
        \textbf{Device} & \textbf{A100 PCIe} & \textbf{RTX~4090} & \textbf{H800 PCIe} \\
        \midrule
        
        Comp. Capability & \makecell{8.0 \\ (Ampere)} & \makecell{8.9 \\ (Ada Lovelace)} & \makecell{9.0 \\ (Hopper)} \\
        \addlinespace 
        
        SMs * cores per SM & 108 * 64 & 128 * 128 & 114 * 128 \\
        Max Clock Rate & 1410 MHz & 2520 MHz & 1755 MHz \\
        Memory Size & 40 GB & 24 GB & 80 GB \\
        Memory Type & HBM2e & GDDR6X & HBM2e \\
        Memory Clock Rate & 1215 MHz & 10501 MHz & 1593 MHz \\
        Memory Bus & 5120-bit & 384-bit & 5120-bit \\
        Memory Bandwidth & 1555 GB/s & 1008 GB/s & 2039 GB/s \\
        L2 Cache & 40 MB & 72 MB & 50 MB \\
        \addlinespace
        
        \makecell[l]{Combined L1 Cache \\\& Shared Memory per SM} & 192 KB & 128 KB & 256 KB \\
        \addlinespace
        
        \# of Tensor Cores & \makecell{432 \\ (3rd Gen.)} & \makecell{512 \\ (4th Gen.)} & \makecell{456 \\ (4th Gen.)} \\
        
        \midrule 
        
        \multicolumn{4}{l}{\textit{Hopper-Specific Features}} \\
        \addlinespace
        
        DPX hardware & No & No & Yes \\
        Distributed Shared Memory & No & No & Yes \\
        Tensor Memory Accelerator & No & No & Yes \\
        
        \bottomrule
    \end{tabular}
\end{table}
\section{Memory Subsystem}
\label{sec:mem}
\begin{figure*}
    \centering
    \includegraphics[width=\linewidth]{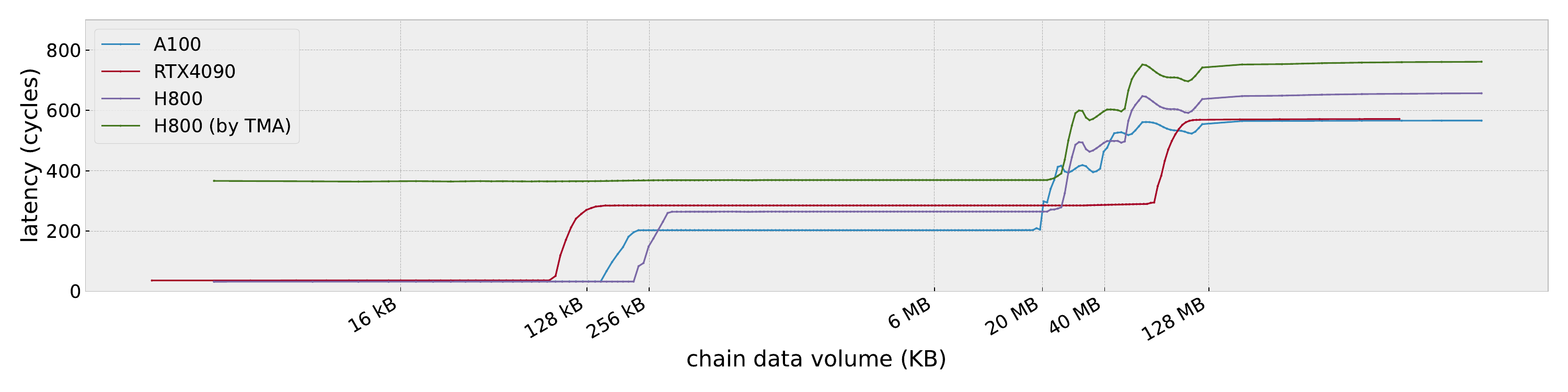}
    \caption{Latency clocks of different memory scopes}
    \label{fig:random-latency}
\end{figure*}
In the Hopper architecture, as shown in Fig. \ref{fig:hopper}, the memory subsystem design closely follows the previous generation, featuring L1 and L2 caches and programmable shared memory. 
Global memory has the highest latency among all GPU memory types but offers the largest capacity. 
In this section, we benchmark the conventional memory hierarchies and focus on two memory performance metrics: latency and throughput. Below we will introduce how we test these two metrics. The newly introduced Tensor Memory Accelerator and  Distributed Shared Memory will be discussed in Sections \ref{sec:tma} and \ref{sec:dsm}, respectively.

\subsection{Latency}
\label{sec:mem_lat}

This section presents latency measurements for the L1 cache, shared memory, L2 cache, and global memory. We find that traditional latency testing methodologies are insufficient for characterizing the performance of partitioned L2 caches. Our fine-grained analysis reveals the specific latency characteristics of this cache architecture.

\noindent\textbf{Traditional P-chase benchmarking.}
We begin by employing the traditional P-chase microbenchmarking method~\cite{Saavedra_Smith_1995,Saavedra_Barrera_1992} with random strides~\cite{gpu_benches} to investigate fundamental memory hierarchy characteristics. This includes examining the range of latencies observed and identifying the array size that triggers the latency inflection points. We read data in 64-byte chunks, which corresponds to the size of a generic address.
For the L1 cache, L2 cache and global memory, we progressively increase the data volume and capture the inflection points in data access latency. In both Ampere\footnote{\url{https://images.nvidia.cn/aem-dam/en-zz/Solutions/data-center/nvidia-ampere-architecture-whitepaper.pdf}} and Hopper\footnote{\url{https://resources.nvidia.com/en-us-data-center-overview/gtc22-whitepaper-hopper}} architectures, the L2 cache is divided into two partitions interconnected by high-speed channels. The GPCs are also partitioned in the same way as the L2 cache. Compared to the commonly used uniform stride in previous work\cite{10579250, wong2010demystifying}, random stride better reflects the existence of two L2 cache partitions.

Fig.~\ref{fig:random-latency} shows the access latencies of different memory levels, which are the average values of more than one million accesses. Before the first inflection point, the access latency reflects L1 cache, around 32-33 clocks for all three devices. Between the first and final inflection points, the latency can be attributed to L2 cache. For RTX 4090, L2 cache latency is approximately 284.8 clocks. At the RTX 4090's third inflection point, the data size roughly equals its L2 cache capacity. As the chain data volume increases, the latency stabilizes around 571 clocks. However, in A100 and H800, we observe two inflection points corresponding to two different latencies for L2 cache. For A100, it is about 202.8 to 408 clocks. In the H800, L2 cache latency is around 264.5 to 502 clocks. 
In the A100 and H800, the third and fourth points are approximately half of the L2 cache size and the full L2 cache size, respectively. 
The stable latency after the fourth point, reflecting global memory latency, is 566 clocks for A100 and 656 clocks for H800. 
Table \ref{tab:mem_lat} concludes the memory access latency at different memory levels.
For shared memory, we initialize the P-chase array with uniform 4-byte strides and then access this array using a thread to measure latency. Since shared memory and the L1 cache are physically the same memory component on GPU hardware, we observe similar latency results, approximately 29$\sim$31 cycles.
When comparing latency across different memory levels, we observe that, on all three devices, the average latency of the L2 cache is approximately 6.5 times higher than that of the L1 cache, while the global memory latency is about 2.1 times higher than that of the L2 cache, with slight variations across devices.

The differing L2 cache latencies observed in the A100 and H800 GPUs arise from their partitioned architectures, which traditional P-chase benchmarking methods cannot fully characterize. To address this limitation, we conduct a subsequent fine-grained analysis.


\begin{table}[ht]
    \centering
    \caption{Latency in clock cycles for different memory scopes using traditional P-chase}
    \label{tab:mem_lat}
    
    \begin{tabular}{l ccc}
        \toprule
        \textbf{Type} & \textbf{RTX4090} & \textbf{A100} & \textbf{H800} \\ 
        \midrule
        
        L1 Cache      & 32.0             & 33.0          & 32.0          \\
        Shared Memory & 30.1             & 29.0          & 29.0          \\
        L2 Cache      & 273.0            & 202.8$\sim$408 & 264.5$\sim$502 \\
        Global Memory & 571              & 566           & 656           \\
        
        \bottomrule
    \end{tabular}
\end{table}

\noindent\textbf{Fine-grained P-chase benchmarking.}
We employ fine-grained P-chase \cite{mei_tpds_2017} to measure the latency of each memory access, revealing patterns in GPU L2 cache and global memory latency. 
Our analysis is based on the \emph{following assumptions}: 1)~Two L2 cache partitions have the same capacity. 
2)~When the size of the data allocated for benchmarking is significantly smaller than half of the total L2 cache capacity, we assume that all memory accesses remain within the cache partition, thus, the result of P-chase does not demonstrate global memory latency. 
3)~Due to the architectural layout of the GPU, there are differences in the physical proximity between streaming multiprocessors (SMs) and L2 cache partitions. 
Accessing an L2 partition closer to an SM results in lower latency compared to accessing a farther partition; 
4)~Any measured latency significantly higher than what can be explained by cache capacity, data size being fully cached, or SM proximity to L2 partitions should be attributed to global memory access due to cache misses. 
The access sequence of fine-grained P-chase aligns with a uniform 32-byte stride. 

On the A100, with an array length of 30MB, four distinct latency groups emerged, corresponding to the four categories in the Table \ref{tab:fg-pchase}. 
In the H800, an 8MB array exhibited two latency types: L2 cache hit near partition (``\textbf{near hit}'') and L2 cache hit far partition (``\textbf{far hit}''). 
With a 40MB array, three latency groups were observed: L2 cache near partition hit, L2 cache near partition miss (``\textbf{near miss}''), and L2 cache far partition miss (``\textbf{far miss}''). 
We apply the K-means clustering algorithm to each device's four data groups, with cluster center results shown in Table~\ref{tab:fg-pchase}. 
We observe that the A100 and H800, with their dual-partitioned L2 caches, exhibit significantly different behavior compared to the RTX 4090, which has an unpartitioned L2 cache. 
Additionally, their global memory access latency is considerably higher than that of the RTX 4090.

We also compare the results between Tables \ref{tab:mem_lat} and \ref{tab:fg-pchase}. In the traditional P-chase benchmarking, when the length of the array is less than half the capacity of the L2 cache (between the first and second inflection points), the latencies for A100 and H800 are 202.8 and 264.5 clocks, respectively, aligning with the L2 cache near hit values in Table~\ref{tab:fg-pchase}. 
Furthermore, the global memory latency in Table \ref{tab:mem_lat} approximates (L2 cache near miss + far miss) / 2 from Table \ref{tab:fg-pchase}. 
This outcome aligns with the probabilistic distribution of data accesses across different L2 cache partitions. 

\begin{table}[ht] 
    \centering
    \caption{Latency in clock cycles for different memory scopes using fine-grained P-chase}
    \label{tab:fg-pchase}
    
    \begin{tabular}{l cc}
        \toprule
        \textbf{Type} & \textbf{A100 PCIe} & \textbf{H800 PCIe} \\
        \midrule
        
        L2 Cache Near Hit  & 208.0 & 258.0 \\
        L2 Cache Far Hit   & 356.6 & 414.1 \\
        L2 Cache Near Miss & 474.9 & 555.5 \\
        L2 Cache Far Miss  & 622.7 & 743.7 \\
        
        \bottomrule
    \end{tabular}
\end{table}

In summary, 
While the traditional P-chase analysis with random strides can partially reveal latency variations arising from accessing different L2 cache partitions~(``\textbf{near hit}'' and ``\textbf{far hit}''), it fails to capture the potential mapping between global memory and these partitions, which could contribute to latency differences observed during L2 cache misses~(``\textbf{near miss}'' and ``\textbf{far miss}'').
However, fine-grained P-chase analysis revealed distinct latency groups on the A100 and H800 GPUs, attributable to their dual-partitioned L2 caches and varying distances between streaming multiprocessors and these partitions. 


\subsection{Throughput}
\label{sec:mem_tp}

In this subsection, we evaluate the throughput of L1 cache, shared memory, L2 cache, and global memory. While memory access modifiers may be unreliable (as demonstrated in the latency tests), we control the volume of accessed data to ensure measurements target the desired cache level.

For the L1 cache test, we also first load the memory into the L1 cache using the \texttt{ca} modifier. Since the L1 cache is exclusive to SM, we issue a block with 1024 threads to access the L1 cache repeatedly. 

Since shared memory can only be accessed within a block (distributed shared memory is not considered in this subsection), we use a block with 1024 threads to access the shared memory repeatedly. 

For the L2 cache test, we first load the memory into the L2 cache using the \texttt{cg} modifier. 
Since the L2 cache is shared by all SMs, the number of blocks we used is twice the number of SMs. 

For the global memory test, we allocate significantly more memory space than the capacity of L2 to bypass the hardware-level cache prefetch mechanism, as detailed in \cite{mei_tpds_2017}. 
In order to reduce the number of memory access instructions, we set up each thread to use vectorized memory access to read \texttt{float4}. Each thread reads 5 times and writes 1 time. The number of blocks we used is four times the number of SMs. The number of blocks has little effect on L2 performance testing, requiring only enough instructions to saturate the hardware units.

For the first three units, data is accessed as both single-precision floating-point values (\texttt{float}) and four-element vectors (\texttt{float4}). While the float type is more common, float4 allows for data transfer with fewer instructions. For global memory accesses, we exclusively employ vectorized (float4) reads and writes to minimize instruction overhead and maximize memory bandwidth utilization \cite{Luitjens2013CUDAPro}.
We record the cycles or time consumed and the amount of data accessed to calculate the throughput of different units.

Table~\ref{tab:mem_bw} shows the memory access throughput at different memory levels. In the cache throughput test, we use different data types for memory access. 
We observe that using vectorized memory access (\texttt{FP32.v4}, equivalent to CUDA's \texttt{float4}) can always achieve better performance. Notably, on the consumer-grade RTX 4090, using FP32 to access L1 cache and shared memory yields only half the throughput achieved with FP32.v4. This performance discrepancy is attributed to memory input/output throttling on the RTX 4090. Utilizing wider data types or increasing the number of blocks can mitigate this bottleneck.

The maximum throughput of L1 cache and shared memory of the three devices are similar because they share the same hardware design. 
However, in terms of L2 cache throughput, the H800 is 2.6 times and 2.2 times higher than the RTX 4090 and A100, respectively.\footnote{Note that for the L1 cache, the amount of data they transfer per clock is almost the same. However, since the order of clock frequency from high to low is RTX4090, H800, and A100, the order of throughput per unit time from high to low is also RTX4090, H800, and A100. The same calculation method also applies to L2 cache.} In the memory throughput test, our results reach 92\%, 90\%, and 91\% of the theoretical performance on RTX4090, A100, and H800 respectively. In the comparison of L2 and Global, the L2 cache throughput of RTX4090, A100, and H800 is 4.67, 2.01, and 4.23 times the global memory throughput, respectively.

\begin{table}[!t] 
    \centering
    \caption{
        Throughput at different memory levels. The unit for L1 cache and shared memory throughput is (byte/clk/SM). The unit for L2 cache throughput is (byte/clk). The unit for global memory throughput is GB/s.
    }
    \label{tab:mem_bw}
    
    \begin{tabular}{l rr rr rr}
        \toprule
        \multirow{2}{*}{\textbf{Type}} & \multicolumn{2}{c}{\textbf{RTX4090}} & \multicolumn{2}{c}{\textbf{A100}} & \multicolumn{2}{c}{\textbf{H800}} \\
        
        \cmidrule(lr){2-3} \cmidrule(lr){4-5} \cmidrule(lr){6-7}
        & \textbf{FP32} & \textbf{FP32.v4} & \textbf{FP32} & \textbf{FP32.v4} & \textbf{FP32} & \textbf{FP32.v4} \\
        \midrule

        L1 Cache     & 63.7   & 121.2  & 99.5   & 106.8  & 125.8  & 124.1  \\
        Shared Memory & 63.7   & 126.5  & 127.9  & 126.2  & 127.4  & 127.9  \\
        L2 Cache     & 1622.2 & 1708.0 & 1853.7 & 2007.9 & 4472.3 & 3942.4 \\
        
        \midrule 

        Global Memory & \multicolumn{2}{c}{929.8} & \multicolumn{2}{c}{1407.2} & \multicolumn{2}{c}{1861.5} \\
        L2 vs. Global & \multicolumn{2}{c}{4.67×} & \multicolumn{2}{c}{2.01×}  & \multicolumn{2}{c}{4.23×}  \\
        \bottomrule
    \end{tabular}
\end{table}
\begin{tcolorbox}[colback=green!5!white, colframe=green!50!black, title=Insight]
1. Hopper's L1 cache and shared memory performance remain consistent with the previous generation.

2. The L2 cache architecture in both Ampere and Hopper features a dual-partition design, where partitions communicate through high-bandwidth interconnection channels. GPCs are similarly organized into two partitions that align with the L2 cache structure, exhibiting optimal performance when accessing proximate L2 cache partitions. 

\end{tcolorbox}

\section{Tensor Memory Accelerator}
\label{sec:tma}

The asynchronous copies introduced by NVIDIA Ampere GPU architecture provide more flexibility for programming. Hopper provides a more sophisticated asynchronous copy engine: the Tensor Memory Accelerator (TMA). As shown in Fig. \ref{fig:hopper}, TMA can be used in two scenarios: (1) transfers between global memory and shared memory in both directions; (2) transfers between shared memory regions of different SMs within a cluster (distributed shared memory).  Ampere's asynchronous copies require all threads to calculate the address of their own memory accesses. However, Hopper's TMA takes care of everything. We can assign one thread to create a copy descriptor before launching the TMA. After that, TMA can automatically handle addressing and data transfer. 

We attempt to expose the address translation efficiency of TMA, which is claimed as the key advantage by the hopper whitepaper. 

\subsection{Latency}

The TMA latency test aligns with the random stride test from Section \ref{sec:mem_lat}\footnote{The only difference is that the minimum load size of TMA is 16 bytes, but we have actually measured that in the latency test, the results obtained with this very small load size are basically the same.}. Our TMA test measures the complete latency, including the issuance of asynchronous TMA instructions, initialization of expected transactions for \texttt{Mbarrier}, and the \texttt{Mbarrier} arrive and wait processes. The results are shown in Fig. \ref{fig:random-latency}. We observe that TMA access has one fewer inflection point compared to regular memory access, suggesting that TMA operations are influenced by L2 cache rather than L1 cache. After the first inflection point of regular memory access, TMA latency trends similarly to global memory access. However, TMA access is approximately 170 cycles higher than regular access, which can be attributed to the overhead of TMA units and synchronization waits.

\subsection{Throughput}

We test the throughput of TMA, as its performance in various scenarios can directly impact our program's efficiency. In this subsection, we load 4GB of data from global memory into shared memory using TMA. We evaluate one-dimensional loads, 1D Tensor, 2D Tensor, and 3D Tensor operations. We focused on the impact of transfer size per TMA instruction and the number of launched CTAs on throughput. For non-tensor (one-dimensional) transfers, TMA instructions can be issued without creating a tensor map. However, for 1D, 2D, and 3D tensor copies, a Copy Descriptor (Tensor Map) as shown in Fig. \ref{fig:hopper} is required, specifying different dimensions. 

For the number of thread blocks, we select 114, 228, 342, and 456, corresponding to 1, 2, 3, and 4 times the number of SMs, respectively. In non-tensor tests, we evaluate load sizes of 1KB, 2KB, 4KB, 8KB, 12KB, and 16KB. For 1D Tensors, the tensor map limits each dimension of the shared memory box to a maximum of 256 elements. Thus, even with \texttt{int64} as the data type, the maximum load is 2KB. We tested 0.5KB, 0.75KB, 1KB, 1.25KB, 1.5KB, and 2KB. In 2D Tensor tests, we use the same sizes as non-tensor tests, with floating-point data types, and shapes of 16×16, 32×16, 32×32, 64×32, 96×32, 64×64. For 3D Tensors, the sizes are also the same as non-tensor tests, with floating-point data types, and shapes of 8×8×4, 8×8×8, 16×8×8, 16×16×8, 16×16×12, 16×16×16. We examine how different combinations of block numbers and TMA load sizes affect global memory utilization.

\begin{figure*}[ht]
    \centering
    \includegraphics[width=\linewidth]{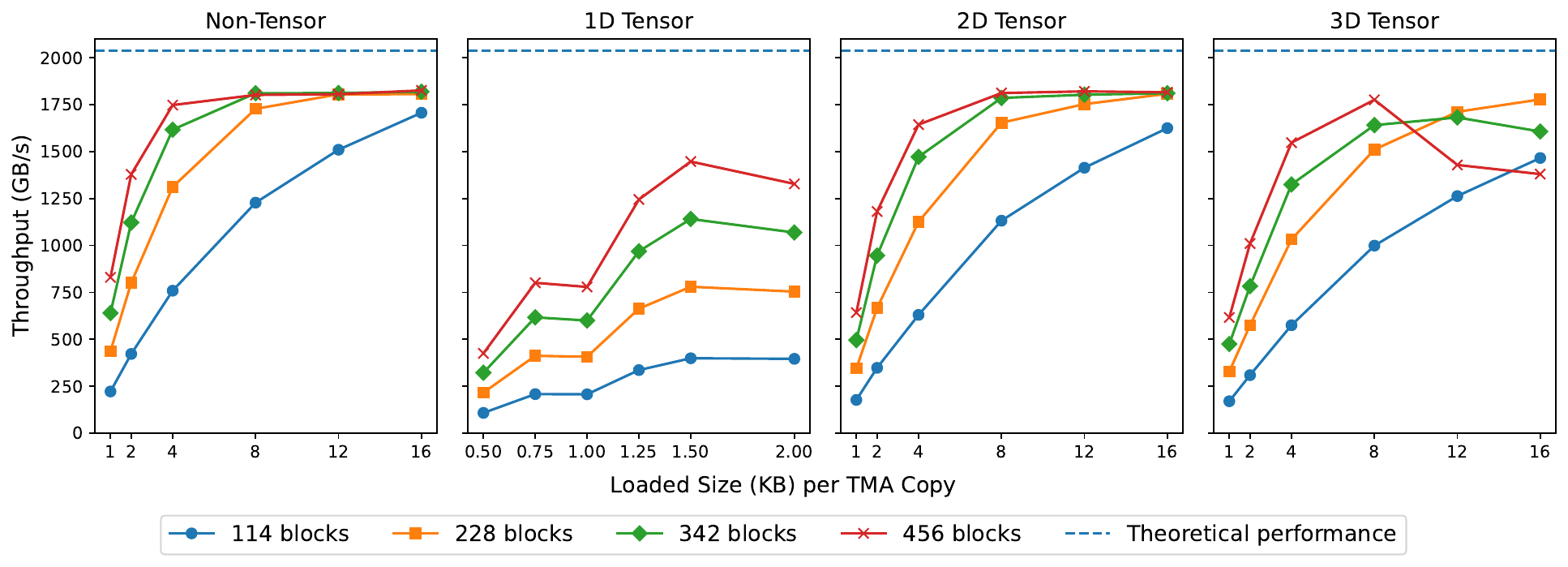} 
    \caption{Global memory throughput of different combinations of block numbers and TMA load sizes}
    \label{fig:tma_tp}
\end{figure*}

The results, shown in Fig. \ref{fig:tma_tp}, indicate that larger load sizes and more thread blocks generally achieve higher memory bandwidth. In non-tensor, 2D tensor, and 3D tensor scenarios, we reach over 1800GB/s, similar to Section \ref{sec:mem_tp}. However, in 1D tensors, the limitations imposed by the tensor map on the shared memory box size prevent further increases in load size. Therefore, for one-dimensional loads, using a non-tensor approach is recommended for better performance. Additionally, in 3D tensors, we observed an unusual phenomenon: when the load size exceeds 8KB, more thread blocks result in lower performance, likely due to the shared memory box constraints.

\begin{figure}[ht]
    \centering
    \includegraphics[width=0.6\linewidth]{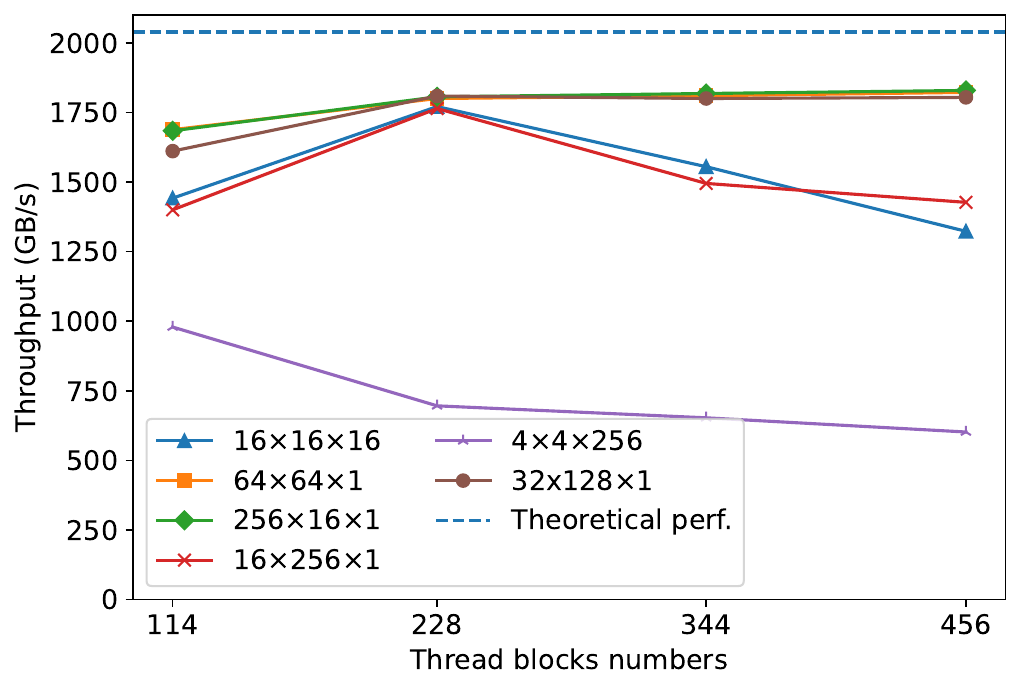} 
    \caption{16KB per TMA load with different shapes}
    \label{fig:tma_3d}
\end{figure}

We compare the performance of different shared memory box shapes with a TMA load size of 16KB and floating-point data type. The configurations tested are 16×16×16, 32x128x1, 64×64×1, 256×16×1, 16×256×1, and 4×4×256. The results, shown in Fig. \ref{fig:tma_3d}, reveal a clear trend: with the same load size, larger x-axis dimensions yield higher throughput without negative effects from increasing thread blocks. However, increasing the y-axis and z-axis dimensions significantly reduces throughput, which may decrease further with more thread blocks. Therefore, when using TMA to handle tensors with dimensions greater than 3D, it's important to carefully select appropriate parameters to achieve optimal performance.

\subsection{Application-level}

For the application-level evaluation of TMA, we select one of the most impactful and widely used applications: General Matrix Multiply (GEMM). GEMM is defined as $\mathbf{D}_{m \times n} = \alpha \mathbf{A}_{m \times k} \times \mathbf{B}_{k \times n} + \beta\mathbf{C}_{m \times n}$. GEMM is extensively applied in fields such as artificial intelligence and scientific computing. GPUs, with their high degree of parallelism, have become the preferred hardware for GEMM computations. To achieve high efficiency on GPUs, GEMM implementations employ tiling strategies, which divide the computational workload into smaller submatrices (tiles) that fit into the memory hierarchy of the GPU. Each tile is assigned to a Cooperative Thread Array (CTA), which consists of threads that collaboratively load data from global memory, perform computations in shared memory, and write results back to global memory. This tiling approach minimizes memory latency and maximizes data reuse, ensuring efficient utilization of GPU resources.

Our implementation is based on the official example provided by CUTLASS\footnote{\url{https://github.com/NVIDIA/cutlass/tree/main/examples/cute/tutorial/hopper}}. To minimize the influence of other factors (e.g., Hopper's more powerful Tensor Cores and its \texttt{wgmma} instructions), this subsection focuses exclusively on comparisons conducted on the Hopper architecture.
In the GEMM implementation without TMA, the programming model follows the same approach as Volta and Ampere, employing pipelining by interleaving stages and instructions. In the TMA-enabled version, explicit producer-consumer synchronization is utilized for purely asynchronous instructions, such as TMA and Tensor Core operations. To maintain high performance for larger matrices, we set the CTA tile sizes for the $M$, $N$, and $K$ dimensions to 128, 256, and 64, respectively. At this point, the TMA load shapes are configured as 
128×256 and 64×256 (with the $\mathbf{B}$ matrix being transposed), which are equivalent to 64×256 and 32×256 in FP32. Based on the instruction-level results, these configurations allow for effective utilization of TMA's performance capabilities. To fully utilize the performance of Tensor Cores (as discussed in Table \ref{tab:Ndiff} of Section \ref{subsec:tc_instruction}, where it is shown that the $N$ dimension needs to be sufficiently large to ensure tensor core performance), we set the $M$, $N$, and $K$ dimensions for Tensor Core instructions to 64, 128, and 16, respectively. Additionally, we configure the pipeline depth to 2. All other parameters remain consistent with the CUTLASS example.

\begin{figure}[ht]
    \centering
    \includegraphics[width=0.6\linewidth]{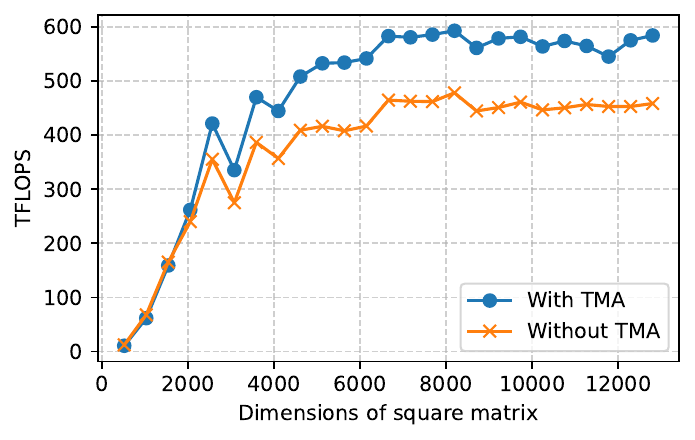} 
    \caption{GEMM Performance Comparison in FP16: With and Without TMA}
    \label{fig:tma_gemm}
\end{figure}

Fig. \ref{fig:tma_gemm} illustrates the performance impact of enabling and disabling TMA in the GEMM application. The horizontal axis represents the dimensions of square matrices $(n\times n)$, while the vertical axis shows the performance in FP16 TFLOPS (Tera Floating-point Operations Per Second).
For smaller matrices $(n < 2000)$, the performance difference between ``With TMA'' (blue line) and ``Without TMA'' (orange line) is minimal, as the smaller workload does not fully leverage GPU’s capabilities. As the matrix size increases $(n \geq 2000)$, the performance with TMA shows a significant advantage, growing faster than the case without TMA. For larger matrices $(n > 7000)$, TMA-enabled performance stabilizes at close to 600 TFLOPS, significantly outperforming the non-TMA case, which levels off at approximately 450 TFLOPS.
This demonstrates that TMA provides substantial benefits for large-scale matrix multiplication by optimizing memory access and data movement, enabling higher computational efficiency and better utilization of hardware resources.

\begin{tcolorbox}[colback=green!5!white, colframe=green!50!black, title=Insight]
1. TMA, as an independent asynchronous memory access unit, enhances the programming flexibility of the Hopper architecture. Utilizing warp specialization can more fully exploit the performance potential of the Hopper architecture.

2. TMA load shapes require careful selection to achieve good performance, ensuring the x-axis dimension is sufficiently large while fully considering memory-compute overlap in the pipeline.
\end{tcolorbox}
\section{Tensor Core}
\label{sec:tc}
\subsection{Tensor Core's Evolution}\label{sec:evo_tc}
Table~\ref{tab:TCS} illustrates the progression of TCs, encompassing enhancements in precision, operand shapes, programming modes, and execution modes. In the initial Volta Architecture, first-generation TCs exclusively supported FP16 as the input data type. 
Subsequent architectures, including Ampere, Ada, and Hopper, introduced support for a broader range of data types such as BF16, TF32, FP64, INT8, INT4, Binary, and more.
The programming of TCs has also seen continuous improvement. Ampere and Ada Lovelace GPUs provide users with the flexibility to utilize either the legacy C-level \texttt{wmma} APIs or PTX-level \texttt{mma} instructions. Notably, the \texttt{wmma} APIs had limitations in fully harnessing TCs' capabilities, whereas \texttt{mma} instructions could leverage advanced sparse matrix multiplication capabilities introduced since Ampere. 
In the case of Hopper GPUs, new warp-group-level \texttt{wgmma} instructions were introduced. Both \texttt{wmma} and \texttt{mma} APIs remain supported in Hopper, but we find that the complete potential of Hopper TCs can only be realized through \texttt{wgmma} instructions.



An \texttt{mma} instruction computes $\mathbf{D}_{m \times n} = \mathbf{A}_{m \times k} \times \mathbf{B}_{k \times n} + \mathbf{C}_{m \times n}$ and is executed synchronously by one CUDA warp (i.e., 32 threads). In contrast, \texttt{wgmma} for Hopper computes $\mathbf{D}_{m \times n} = \mathbf{A}_{m \times k} \times \mathbf{B}_{k \times n} + \mathbf{D}_{m \times n}$ and is executed asynchronously by one CUDA warp group (i.e., four CUDA warps). Here, $\mathbf{A}_{m \times k}$ denotes a matrix $\mathbf{A}$ with $m$ rows and $k$ columns, and similarly for the other matrices. 
The matrix shapes for \texttt{mma} instructions can be $m16n8k16$ or $m16n8k8$, while \texttt{wgmma} supports $m64nNk16$, where $N$ can be 16, 32, 64, 128, 256, and so on~\cite{nvidia-ptx-doc}. 
Notably, \texttt{wgmma} has the advantage of directly loading matrices $A$ and $B$ from shared memory, unlike \texttt{mma}, which requires storing all matrices in the register file before execution. 
We use the term ``SS" to denote the \texttt{wgmma} instruction that loads both $A$ and $B$ from shared memory, while ``RS" is used for the instruction that loads $A$ from the register file. 
Additionally, \texttt{wgmma} offers support for certain useful arguments not required for \texttt{mma}. 
Further details are provided in \cite{nvidia-ptx-doc}. Notably, wgmma instructions are specific to the Hopper architecture, whereas mma instructions offer backward and forward compatibility across generations.


\begin{table}[!ht]
\caption{Properties of the latest generations of Tensor Cores}
\label{tab:TCS}
\addtolength{\tabcolsep}{-2.5pt}
\centering
\begin{tabular}{|l|l|l|l|}
\hline
\textbf{Arch} & \textbf{Precision} & \textbf{Programmability} & \textbf{Mode} \\ \hline
Ampere & \begin{tabular}[c]{@{}l@{}}FP16,BF16,\\ TF32,FP64,\\ INT8,INT4,Binary\end{tabular} & \begin{tabular}[c]{@{}l@{}}C: wmma\\ PTX: mma, mma.sp\end{tabular} & Sync \\ \hline
Ada & \begin{tabular}[c]{@{}l@{}}FP16,BF16,FP8,\\ TF32,FP64,\\ INT8,INT4,Binary\end{tabular} & \begin{tabular}[c]{@{}l@{}}C: wmma\\ PTX:mma, mma.sp\end{tabular} & Sync \\ \hline
\multirow{2}{*}{Hopper} & \multirow{2}{*}{\begin{tabular}[c]{@{}l@{}}FP16,BF16,FP8,TF32,\\ FP64,INT8,Binary\end{tabular}} & \begin{tabular}[c]{@{}l@{}}C: wmma \\ PTX: mma, mma.sp\end{tabular} & Sync \\ \cline{3-4} 
 &  & PTX: wgmma, wgmma.sp & ASync \\ \hline
\end{tabular}
\end{table}

\subsection{Instruction-level}
\label{subsec:tc_instruction}

We conduct micro-benchmarking of TCs at the PTX level, as it strikes a suitable balance between granularity and complexity. Additionally, we disassemble PTX instructions to SASS codes to achieve a deeper understanding of the operations. SASS analysis is provided in the supplementary materials, which are available online.

We focus on assessing two critical metrics: latency and throughput. Latency signifies the elapsed time, measured in clock cycles, starting from the initiation of instruction issuance to the execution pipeline and concluding when the results become accessible for subsequent usage. This measurement is specifically labeled as ``completion latency." To elaborate further, we issue a single synchronous TC instruction (i.e., \texttt{mma}) using one CUDA warp per SM, whereas one asynchronous TC instruction (i.e., \texttt{wgmma}) is issued utilizing four CUDA warps (comprising a warp group) on one SM. We execute the instruction 1024 times within a CUDA kernel. Throughput is quantified as $\mathit{Total\_{OPS}}/\mathit{Duration}$, where $\mathit{OPS}$ represents multiplication or addition operations. 
It's important to emphasize that, unlike the approach described in \cite{sun_tpds_2023}, we abstain from utilizing total clock cycles to compute throughput due to potential variations in GPU frequencies during the execution of different TC instructions.

\noindent\textbf{\texttt{mma} results.} Table~\ref{tab:syncwgmma} provides an overview of the latency and throughput measurements for \texttt{mma} instructions across A100, RTX4090, and H800 Tensor Cores GPUs. Note that the sparse shapes in the table represent compressed shapes. In other words, the $k$ of the actual instruction modifier is twice that in the table.

\begin{table*}[!ht]
    \centering 
    \caption{
        Different dense and sparse \texttt{mma} instructions on A100, RTX4090 and H800 Tensor Cores. Latency (LAT) is measured in clock cycles. Throughput is measured in TFLOPS or TOPS/s. Peak performance (A100): FP16 (312 TFLOPS); TF32 (156 TFLOPS); INT8 (624 TOPS). Peak performance (RTX4090): FP16 (330.3 TFLOPS); TF32 (82.6 TFLOPS); INT8 (660.6 TOPS). Peak performance (H800): FP16 (756.5 TFLOPS); TF32 (378 TFLOPS); INT8 (1513 TOPS).
    }
    \label{tab:syncwgmma}

    \begin{tabular}{lll rr rr rr}
        \toprule
        \multirow{3}{*}{\textbf{A/B}} & \multirow{3}{*}{\textbf{C/D}} & \multirow{3}{*}{\textbf{Shape}} & \multicolumn{6}{c}{\textbf{LAT/Throughput}} \\
        
        \cmidrule(lr){4-9}
        & & & \multicolumn{2}{c}{\textbf{A100}} & \multicolumn{2}{c}{\textbf{RTX4090}} & \multicolumn{2}{c}{\textbf{H800}} \\
        
        \cmidrule(lr){4-5} \cmidrule(lr){6-7} \cmidrule(lr){8-9}
        & & & \textbf{Dense} & \textbf{Sparse} & \textbf{Dense} & \textbf{Sparse} & \textbf{Dense} & \textbf{Sparse} \\
        \midrule

        FP16 & FP16 & m16n8k8   & 17.7/310.0 & 17.3/408.4 & 17.7/355.3 & 17.3/713.2 & 16.0/368.6 & 16.0/493.8 \\
        FP16 & FP16 & m16n8k16  & 24.6/310.6 & 24.5/622.8 & 24.6/357.6 & 24.5/711.8 & 24.1/494.4 & 24.0/722.8 \\
        FP16 & FP32 & m16n8k8   & 17.5/299.6 & 18.0/394.1 & 18.8/177.8 & 18.8/357.4 & 16.0/363.7 & 16.0/488.7 \\
        FP16 & FP32 & m16n8k16  & 26.0/303.4 & 24.5/603.3 & 33.0/178.9 & 33.0/356.0 & 24.1/490.7 & 24.0/721.8 \\
        TF32 & FP32 & m16n8k4   & 17.8/149.5 & 18.2/196.8 & 19.2/89.0  & 19.0/178.0 & 16.5/180.6 & 16.4/240.7 \\
        TF32 & FP32 & m16n8k8   & 26.3/151.5 & 26.7/301.5 & 33.4/89.0  & 33.3/178.7 & 24.5/246.4 & 24.4/363.3 \\
        INT8 & INT32 & m16n8k16 & 17.6/594.8 & 18.0/788.5 & 17.3/707.6 & 17.3/1412  & 16.1/730.3 & 16.1/970.0 \\
        INT8 & INT32 & m16n8k32 & 26.0/607.6 & 26.6/1210  & 24.5/711.7 & 24.6/1423  & 24.0/977.9 & 24.2/1435  \\
        \bottomrule
    \end{tabular}
\end{table*}

For A100 and H800, the same-precision \texttt{mma} instructions with the larger shapes commonly achieve better throughputs. But this phenomenon disappears on RTX4090. Sparse and dense \texttt{mma} instructions exhibit equivalent latency, with sparse \texttt{mma} instructions achieving higher throughputs. On the RTX4090, sparse \texttt{mma} instructions can achieve up to double the throughput compared to their corresponding dense counterparts, aligning with the speedup claims stated in the vendor's documentation. However, for the A100, only the sparse \texttt{mma} instructions with larger shapes can realize the theoretical speedups. In the case of the H800, sparse \texttt{mma} instructions can only achieve an average speedup of 1.42 times over the dense ones. 
This highlights that on Hopper Tensor Cores, sparse \texttt{mma} instructions may not fully harness the capabilities of the sparse tensor cores.

The achieved throughput on A100 exceeds 95\% of their theoretical peak performance. 
The achieved throughput of RTX4090 is higher than the official theoretical peak performance. This is because our RTX4090 runs at a higher frequency (2710 MHz) than the officially announced boost frequency. 
However, on Hopper Tensor Cores, \texttt{mma} instructions can only attain an average of 62.9\% of the theoretical peak performance.
It indicates that the \texttt{mma} instructions should be used carefully as they cannot fully utilize the tensor cores' potential in some cases.

\noindent\textbf{\texttt{wgmma} results.} As a set of warp-group-level Tensor Core instructions designed specifically for Hopper GPUs, \texttt{wgmma} instructions are the pioneering instructions to be executed asynchronously. 
Table~\ref{tab:densewgmma} show the measured latency and throughput of dense instructions. 
When initializing matrices with zeros, we achieve throughputs exceeding 95\% of the theoretical peak performance. We observe a decrease in Tensor Core performance when initializing matrices with random values, especially pronounced when utilizing FP16 as the computation type and FP32 for accumulation. 
This phenomenon is primarily attributed to the power consumption nearing the 350~W power limit of the H800-PCIe, subsequently causing a reduction in frequency. Users working with Tensor Cores on the H800-PCIe GPU should take into full consideration the power constraints when performing computations.

   \begin{table*}[!ht]
    \centering 
    \caption{
        Variations in Dense \texttt{wgmma} Instructions for H800 Tensor Cores. Latency (LAT) is quantified in clock cycles, while throughput is expressed in TFLOPS or TOPS. The peak throughputs for FP16, TF32, FP8, and INT8 are 756.5, 373, 1513, and 1513, correspondingly. ``Zero" or ``Rand" signifies that all matrices are initialized with either zero or randomly generated values. ``SS" implies that both matrix A and B are stored in shared memory, while ``RS'' signifies that matrix A is stored in the register file, whereas B is stored in shared memory. ``Rand" has the same latency as ``Zero".
    }
    \label{tab:densewgmma}
    

    \begin{tabular}{l l l c c c c}
        \toprule 
        \textbf{A/B} & \textbf{C/D} & \textbf{Instruction} &
        \makecell{\textbf{LAT/Throughput} \\ \textbf{(SS,Zero)}} &
        \makecell{\textbf{LAT/Throughput} \\ \textbf{(RS,Zero)}} &
        \makecell{\textbf{Throughput} \\ \textbf{(SS,Rand)}} &
        \makecell{\textbf{Throughput} \\ \textbf{(RS,Rand)}} \\
        \midrule 

        FP16 & FP16  & m64n256k16 & 128.0/729.3  & 128.0/729.2  & 704.5  & 703.7  \\
        FP16 & FP32  & m64n256k16 & 128.0/728.5  & 128.0/731.9  & 665.4  & 667.5  \\
        TF32 & FP32  & m64n256k8  & 128.0/364.4  & 128.0/364.6  & 357.1  & 357.3  \\
        FP8  & FP16  & m64n256k32 & 128.0/1448.4 & 128.0/1448.0 & 1439.2 & 1440.3 \\
        FP8  & FP32  & m64n256k32 & 128.0/1447.5 & 128.0/1455.0 & 1417.2 & 1419.8 \\
        INT8 & INT32 & m64n256k32 & 128.0/1448.7 & 128.0/1447.9 & 1442.3 & 1442.2 \\
        \bottomrule 
    \end{tabular}
\end{table*}

In the case of dense \texttt{wgmma}, with $N$ set to 128, we observe that the latency for all data types corresponding to the instructions is 128.0. Interestingly, under both ``RS'' and ``SS'' modes, the latency and throughput for the same instruction remain relatively consistent. 
This is due to the effective hiding of shared memory latency through high computational workload and asynchronous operations. 

In the context of sparse \texttt{wgmma} instructions, the latencies for ``RS'' and ``SS'' modes are 128.0 and 144.0, respectively. Additionally, we observe that in the "SS" mode (Rand), sparse \texttt{wgmma} achieves approximately 1.8x performance improvement compared to dense \texttt{wgmma}. In the "RS" mode (Rand), the performance improvement exceeds 1.9x. However, the performance of sparse \texttt{wgmma} in the "SS" mode is lower than in the "RS" mode, which is notably different from the behavior of dense \texttt{wgmma}.
We find that in sparse \texttt{wgmma}, the ``SS'' mode retrieves data from the shared memory of size $m \times k$ and performs a 2:4 sparse pruning based on metadata during the execution of the sparse \texttt{wgmma} instruction. 
In contrast, the ``RS'' mode directly accesses data from the pruned register file of size $m \times k / 2$. The high shared memory access demand (twice as much) may lead to latency that cannot be effectively concealed by Tensor Core computation, resulting in sparse \texttt{wgmma} instructions in ``SS'' modes failing to achieve the expected peak performance.

\noindent\textbf{\texttt{wgmma} results with different $N$ values.} We conduct tests using the example of \texttt{wgmma.m64nNk16.f32.f16.f16}, varying the value of $N$, and the results are presented in Table~\ref{tab:Ndiff}. When $N$ is greater than or equal to 64, all \texttt{wgmma} instructions can achieve throughputs that closely approach peak performance.
However, when $N$ is less than 64, the achieved throughput decreases, and the ``SS'' mode of instructions exhibits higher latency than the ``RS'' mode, while the achieved throughputs are lower than those of the ``RS'' mode. 
As $N$ decreases, the computational density of \texttt{wgmma} instructions gradually diminishes, making it challenging to conceal the latency associated with shared memory access, leading to the aforementioned phenomena. 
Therefore, when utilizing wgmma instructions, it is advisable to opt for larger values of $N$ ($>=64$) whenever possible to attain superior performance.

\begin{table*}[!ht]
    \centering
    \caption{
        Different \texttt{wgmma} instructions with different $N$ values on H800 tensor cores. The definitions of LAT and throughput can be found in the caption of Table \ref{tab:densewgmma}. ``Rand" has the same latency as ``Zero".
    }
    \label{tab:Ndiff}

    \begin{tabular}{l rrrr rrrr}
        \toprule
        \multirow{3}{*}{\textbf{N}} & \multicolumn{4}{c}{\textbf{Dense}} & \multicolumn{4}{c}{\textbf{Sparse}} \\
        
        \cmidrule(lr){2-5} \cmidrule(lr){6-9}
        
        & \makecell{LAT/Tput \\ (SS,Zero)} & \makecell{LAT/Tput \\ (RS,Zero)} & \makecell{Tput \\ (SS,Rand)} & \makecell{Tput \\ (RS,Rand)}
        & \makecell{LAT/Tput \\ (SS,Zero)} & \makecell{LAT/Tput \\ (RS,Zero)} & \makecell{Tput \\ (SS,Rand)} & \makecell{Tput \\ (RS,Rand)} \\
        \midrule

        256 & 128.0/728.5 & 128.0/731.9 & 665.4 & 667.5 & 144.0/1312.3 & 128.0/1476.2 & 1194.3 & 1277.5 \\
        128 & 64.0/728.5  & 64.0/725.4  & 659.8 & 661.7 & 80.0/1176.4  & 64.0/1463.3  & 1109.6 & 1270.5 \\
        64  & 32.0/719.6  & 32.0/719.7  & 648.3 & 649.9 & 48.0/977.4   & 32.0/1450.1  & 969.9  & 1263.4 \\
        32  & 24.0/477.3  & 16.0/710.3  & 471.5 & 634.4 & 32.0/727.1   & 18.0/1272.4  & 723.4  & 1135.7 \\
        16  & 20.0/287.0  & 13.0/434.2  & 283.5 & 426.2 & 24.0/482.3   & 18.0/638.6   & 479.8  & 636.3  \\
        8   & 18.0/158.2  & 13.0/216.7  & 157.6 & 215.2 & 20.0/289.0   & 16.0/359.4   & 286.1  & 356.7  \\
        \bottomrule
    \end{tabular}
\end{table*}

\noindent\textbf{Energy efficiency.} 
Although Tensor Cores have impressive performance, their energy efficiency should also be considered. The training of GPT-3 consumes approximately 1,287 megawatt-hours of electricity in a single training run\cite{maslej2023artificial}. Since tensor core instructions constitute the primary computational workload in AI training, analyzing their energy consumption is essential for developing green AI solutions\cite{tang2019impact, wang2020benchmarking, nabavinejad2022coordinated}. Therefore, we conduct comprehensive energy efficiency analysis to understand the power characteristics of different tensor core instructions.

In our benchmarks, we define energy efficiency as performance per watt, measured in TFLOPS(TOPS)/watt.
Both \texttt{mma} and \texttt{wgmma} are considered in this work.
Our methodology involves repeatedly executing either \texttt{mma} or \texttt{wgmma} instructions across all SMs, with a minimum of two billion repetitions. We utilize the tools provided in \cite{tpds2020_gpudvfs} to monitor both frequency and performance. The energy efficiency of each instruction is determined by calculating the ratio of performance to power consumption once steady-state operation is achieved during these repetitive executions.

For \texttt{mma} instructions, we test the largest operational shape in Table \ref{tab:syncwgmma}. The energy efficiency of \texttt{mma} instructions is shown in Table \ref{tab:energy}. 
In terms of dense instructions, the average energy efficiency of H800 is 1.60 times and 1.69 times that of A100 and RTX4090 respectively. In terms of sparse instructions, the average energy efficiency of H800 is 1.33 times and 1.39 times that of A100 and RTX4090 respectively. 
We find that the H800 has significantly higher energy efficiency.

\begin{table}[htbp]
    \centering
    \caption{
        Power consumption and energy efficiency of maximum shape under \texttt{mma} instructions. Energy is measured in Watts. Efficiency is measured in TFLOPS(TOPS)/Watt.
    }
    \label{tab:energy}

    \begin{tabular}{lll cccccc}
        \toprule
        \multirow{2}{*}{\textbf{A/B}} & \multirow{2}{*}{\textbf{C/D}} & \multirow{2}{*}{\textbf{Type}} & \multicolumn{2}{c}{\textbf{A100}} & \multicolumn{2}{c}{\textbf{H800}} & \multicolumn{2}{c}{\textbf{RTX4090}} \\
        \cmidrule(lr){4-5} \cmidrule(lr){6-7} \cmidrule(lr){8-9} 
        & & & \textbf{Energy} & \textbf{Efficiency} & \textbf{Energy} & \textbf{Efficiency} & \textbf{Energy} & \textbf{Efficiency} \\
        \midrule
        
        \multirow{2}{*}{FP16} & \multirow{2}{*}{FP16} & Dense & 173.4 & 1.79 & 188.6 & 2.62 & 189.1 & 1.89 \\
        & & Sparse & 198.8 & 3.13 & 187.2 & 3.86 & 214.0 & 3.33 \\
        \midrule 

        \multirow{2}{*}{FP16} & \multirow{2}{*}{FP32} & Dense & 188.5 & 1.61 & 196.7 & 2.49 & 154.1 & 1.16 \\
        & & Sparse & 216.1 & 2.79 & 194.9 & 3.70 & 165.9 & 2.15 \\
        \midrule 

        \multirow{2}{*}{TF32} & \multirow{2}{*}{FP32} & Dense & 214.7 & 0.71 & 254.9 & 0.97 & 174.3 & 0.51 \\
        & & Sparse & 235.7 & 1.28 & 232.5 & 1.56 & 187.9 & 0.95 \\
        \midrule 

        \multirow{2}{*}{INT8} & \multirow{2}{*}{INT32} & Dense & 178.4 & 3.41 & 165.3 & 5.92 & 201.4 & 3.53 \\
        & & Sparse & 193.9 & 6.24 & 163.3 & 8.79 & 219.8 & 6.47 \\
        
        \bottomrule
    \end{tabular}
\end{table}

For \texttt{wgmma} instructions, we focus on both the frequency and efficiency when these instructions are executed continuously. The instruction formats are consistent with those described in Table \ref{tab:densewgmma}. 
The results are presented in Table \ref{tab:wgmma-eff}. 
It is observed that, during continuous execution of \texttt{wgmma} instructions, the core frequency drops below 1620~MHz specified in the whitepaper, as it reaches the 350-watt power limit. 
Furthermore, the initialization of inputs significantly impacts both power consumption and frequency. 
When inputs are set to zero, power consumption remains below 200 watts. However, with random inputs, the power consumption rapidly hits the 350-watt threshold. 
Consequently, the performance results in Table~\ref{tab:wgmma-eff} are inferior to those in Table \ref{tab:densewgmma}. 
Notably, sparse instructions cause a substantial reduction in frequency, explaining why their performance is not twice that of dense instructions. 
Regarding energy efficiency, \texttt{wgmma} dense and sparse instructions achieve average 0.67 times and 0.78 times the efficiency of \texttt{mma} dense and sparse instructions, respectively. 
Compared to different architectures, the energy efficiency of Hopper \texttt{wgmma} is on average 1.05 times and 1.10 times that of Ampere \texttt{mma} and Ada \texttt{mma} instructions, respectively. 

Based on these results, we conclude that \texttt{wgmma} instructions can fully utilize the GPU's potential to achieve maximum performance, but they do not exhibit good energy efficiency. 
If extreme performance is desired, \texttt{wgmma} instructions should be chosen. However, if greater emphasis is placed on energy efficiency and compatibility, \texttt{mma} instructions are the better choice.

\begin{table}[htbp]
    \centering
    \caption{
        Power consumption and energy efficiency of maximum shape under \texttt{wgmma} instructions. Frequency is measured in MHz. Performance is measured in TFLOPS (TOPS). Efficiency is measured in TFLOPS (TOPS)/watt.
    }
    \label{tab:wgmma-eff}

    \begin{tabular}{lll rrrr}
        \toprule
        \textbf{A/B} & \textbf{C/D} & \textbf{Type} & \makecell{Freq/Perf \\ (SS,Zero)} & \makecell{Freq/Perf \\ (RS,Zero)} & \makecell{Freq/Perf/Effi \\ (SS,Rand)} & \makecell{Freq/Perf/Effi \\ (RS,Rand)} \\
        \midrule
        
        \multirow{2}{*}{FP16} & \multirow{2}{*}{FP16} & Dense & 1560/729 & 1590/743 & 1275/595/1.70 & 1275/600/1.71 \\
        & & Sparse & 1440/1192 & 1380/1290 & 1200/997/2.85 & 1185/1102/3.15 \\
        \midrule 

        \multirow{2}{*}{FP16} & \multirow{2}{*}{FP32} & Dense & 1485/688 & 1530/714 & 1260/583/1.67 & 1290/594/1.70 \\
        & & Sparse & 1380/1142 & 1320/1234 & 1230/998/2.85 & 1170/1081/3.09 \\
        \midrule 

        \multirow{2}{*}{TF32} & \multirow{2}{*}{FP32} & Dense & 1755/383 & 1755/384 & 1425/322/0.92 & 1425/327/0.93 \\
        & & Sparse & 1560/644 & 1515/707 & 1320/553/1.58 & 1245/581/1.66 \\
        \midrule 

        \multirow{2}{*}{FP8} & \multirow{2}{*}{FP16} & Dense & 1590/1484 & 1620/1512 & 1305/1213/3.47 & 1320/1239/3.54 \\
        & & Sparse & 1455/2414 & 1395/2605 & 1275/2122/6.06 & 1200/2256/6.45 \\
        \midrule 

        \multirow{2}{*}{FP8} & \multirow{2}{*}{FP32} & Dense & 1500/1401 & 1530/1426 & 1290/1202/3.43 & 1260/1180/3.37 \\
        & & Sparse & 1395/2316 & 1335/2520 & 1245/2052/5.86 & 1185/2205/6.30 \\
        \midrule 

        \multirow{2}{*}{INT8} & \multirow{2}{*}{INT32} & Dense & 1755/1533 & 1755/1534 & 1380/1285/3.67 & 1395/1306/3.73 \\
        & & Sparse & 1755/2580 & 1725/2899 & 1290/2144/6.13 & 1215/2275/6.50 \\
        
        \bottomrule
    \end{tabular}
\end{table}

\subsection{Library-level}

The Transformer Engine (TE)~\cite{te} is a library specifically designed to accelerate Transformer models~\cite{vaswani2017attention}, following the introduction of the Hopper architecture. 
It is capable of leveraging the FP8 precision offered by both the Hopper and Ada architectures. 
In particular, it provides a variety of optimized modules for Transformer layers that can be utilized within the widely-used deep learning framework, PyTorch~\cite{pytorch}. 
In this subsection, we benchmark two modules of the Transformer Engine to explore their potential as key components in large language models.

\subsubsection{Linear Layer}
In the Transformer architecture, most of the computational overhead comes from the linear layers, specifically matrix multiplication while the Transformer Engine provides the \texttt{te.Linear} implementation to perform matrix multiplication with higher throughput on FP8 Tensor Cores.
When employing the Transformer Engine with \texttt{te.Linear} for matrix multiplication in FP8 precision, TE converts both the input and weights in the linear layer to FP8. 
This conversion process involves data transformation and quantization operations. For example, as the dynamic range of FP8 may not encompass the maximum value of the input tensor, TE identifies the maximum absolute value of the input data as the scaling factor. It then adjusts the input data to fit the representation range of FP8 using $\mathit{inp_{fp8}} = \mathit{inp_{fp16}} / \mathit{scale}$, followed by matrix multiplication in FP8 Tensor Core $\mathit{out_{fp8}} = \mathit{inp_{fp8}} \times \mathit{w_{fp8}}$. 
Finally, it scales the result with $\mathit{out_{fp16}} = \mathit{out_{fp8}} \times \mathit{scale}$. 
This operation would introduce some overhead.


When performing FP8 matrix multiplication using \texttt{te.Linear}, kernel execution time significantly increases with matrix size. For smaller matrices (N=4096), kernel time accounts for only 25.3\% of the execution, while for larger matrices (N=16384), it dominates at 84.7\%. To evaluate Tensor Engine (TE) optimization for linear layers, we measure the throughput (GFLOPS) of \texttt{te.Linear} using square matrices $\mathbf{D}_{n \times n} = \mathbf{A}_{n \times n} \times \mathbf{B}_{n \times n}$, effectively showcasing its performance characteristics.

    \begin{figure}[htbp]
        \centering
        \includegraphics[width=0.55\linewidth]{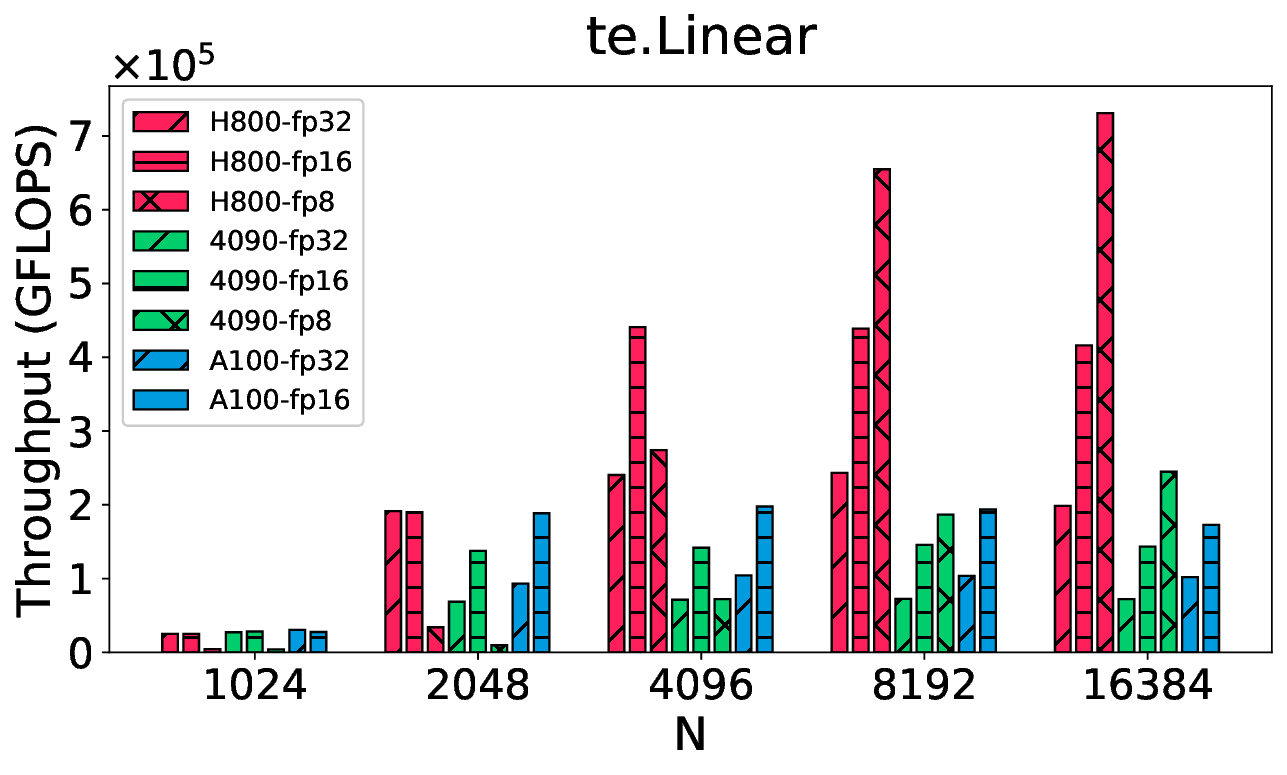} 
        \caption{Comparison of throughput for matrix multiplication with two same-size matrices $\mathbf{D}_{n \times n} = \mathbf{A}_{n \times n} \times \mathbf{B}_{n \times n}$ in different hardware configurations and data types using \texttt{te.Linear}.}
        \label{fig:linear-all}
    \end{figure}
    
We extensively assess the Linear performance across diverse shapes, data types, and hardware setups (Fig. \ref{fig:linear-all}). Leveraging the Transformer Engine, we expedite matrix multiplications for the Linear layer using FP8 Tensor Cores. Our findings reveal an increase in GPU utilization and throughput with larger matrix sizes. FP8 performance is influenced by the overhead from data format conversion and quantization operators. For smaller matrix sizes, FP8 throughput is lower compared to FP16 or FP32. However, with $N$=8192, FP8's performance gains become evident. When $N$=16384, H800 and 4090 utilizing FP8 achieve almost twice the throughput of FP16. This underscores FP8's high throughput potential but underlines the need for specific conditions to attain optimal computing density.


\subsubsection{Transformer Layer}
The TE capitalizes on the efficiency improvements provided by FP8 through specific operator fusion optimizations for transformer layer structures. 
For example, \texttt{te.LayerNormMLP} combines layernorm and MLP within the transformer structure, allowing data transmission between layernorm and the subsequent MLP layer to adopt the FP8 format. This approach not only eliminates data format conversion overhead but also effectively leverages FP8 memory transfer advantages.

TE offers a \texttt{te.TransformerLayer} module that encompasses all operator optimizations for transformer layer structures, facilitating the implementation of various Large Language Model (LLM) structures by adjusting its parameters. However, some operators, such as \texttt{Softmax} and \texttt{GeLU}, have not been quantized to FP8 by TE, resulting in significant data format conversion overhead. Additionally, the DotProductAttention operator, implemented with flash-attention \cite{flashattention}, does not utilize FP8 Tensor Cores.

The computational overhead of the transformer layer's linear layer primarily depends on the hidden size, raising the question of which hidden state (dimension of embedding) will yield better performance for TE with FP8 compared to FP16. 
We investigate this by examining the open-source LLM, Llama~\cite{llama1, llama2}, modifying the activation function to SwiGLU~\cite{swiglu} and normalization to RMSNorm~\cite{rmsnorm}. We set layer structure parameters based on the hidden state's size, with hidden states 4096, 5120, and 8192 corresponding to Llama configurations 7b, 13b, and 70b, respectively.


\begin{table}[!ht]
    \centering
    \caption{Parameter settings of \texttt{te.TransformerLayer} for various \texttt{hidden\_size} values}
    \label{tab:telayer}
    
    \begin{tabular}{l rrrrr}
        \toprule
        \texttt{hidden\_size} & 1024 & 2048 & 4096 & 5120 & 8192 \\
        \midrule 
        \texttt{ffn\_hidden\_size} & 2816 & 5632 & 11008 & 13824 & 22016 \\
        \texttt{num\_attention\_heads} & 8 & 16 & 32 & 40 & 64 \\
        \bottomrule
    \end{tabular}
\end{table}

We fixed the input as (4, 512, $hidden\_size$), where 4 is the $batch\_size$, 512 is the $sequence\_length$, and the attention mask is set to None. We then calculated the latency (ms) required for encoding a single layer once, focusing on the encoding task for a single layer.

    \begin{figure}[htbp]
        \centering
        \includegraphics[width=0.55\linewidth]{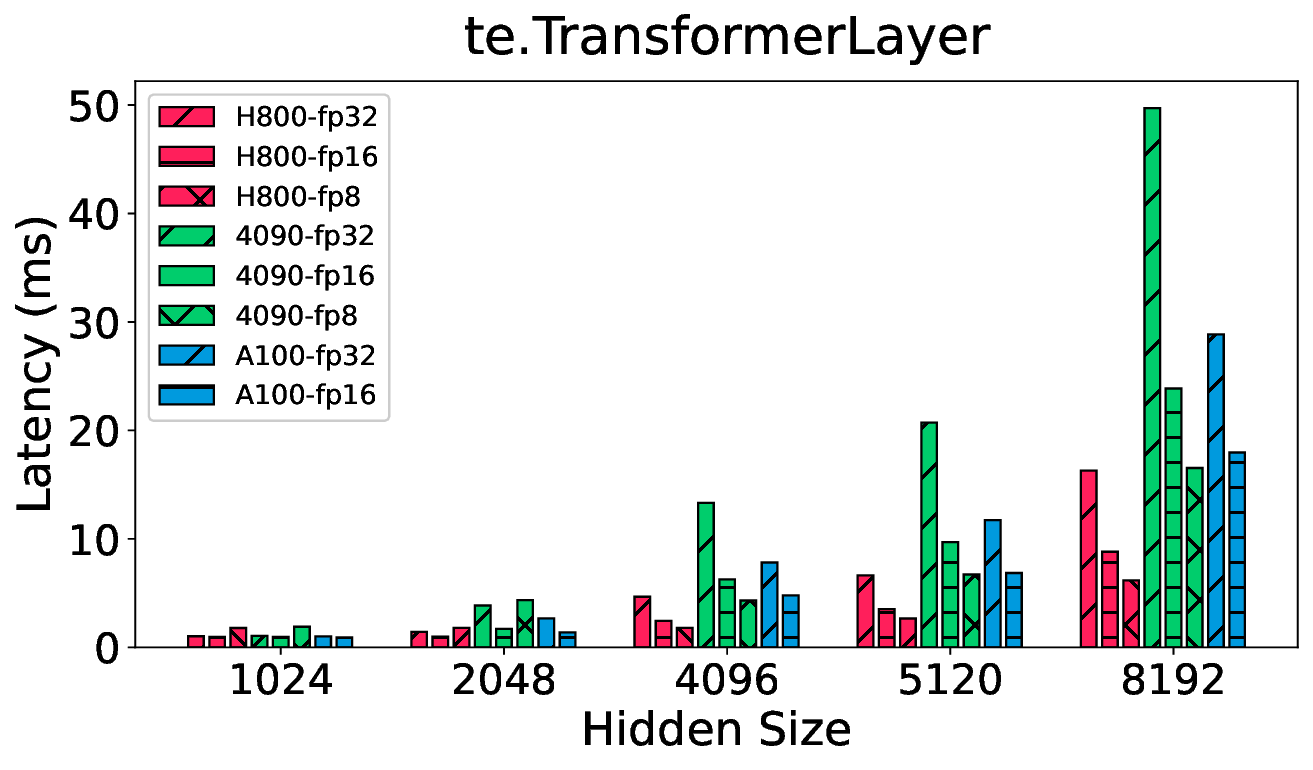} 
        \caption{Comparison of latency for the same input text in different hardware configurations and data types using \texttt{te.TransformerLayer}.}
        \label{fig:TeLayer-all}
    \end{figure}

The Transformer Engine condenses the entire Transformer Layer structure into \texttt{te.TransformerLayer}. Fig. \ref{fig:TeLayer-all} illustrates the latency for the same input text, offering a performance comparison across various hardware setups and data types with \texttt{te.TransformerLayer}. As computational density increases, the advantage of H800 in computation becomes evident. Notably, FP16 shows nearly twice the speed compared to FP32. FP8 outperforms FP16 for hidden\_size$>$4096 but does not achieve double FP16 performance. This is because some modules within the Transformer Layer still do not utilize FP8 precision for calculations and data movement.

\subsection{Application-level}

Currently, the Transformer Engine has not provided optimal support for mainstream decode-only casual language models. In order to test the inference performance of the Transformer Engine on this type of model like, Llama \cite{llama1}, we replaced the \texttt{nn.Linear} and \texttt{RMSNorm} in the original model structure with \texttt{te.Linear} and \texttt{te.RMSNorm}, respectively, to ensure that most modules in the model utilize the Transformer Engine.

To evaluate the effectiveness of TE in generating text for Llama, we follow \cite{vllm} using the ShareGPT dataset as input for the LLM. The ShareGPT dataset comprises conversations between users and ChatGPT \cite{chatgpt}, which have been shared by the users. We tokenize these datasets and, based on their input and output lengths, generate synthesized client requests.

In order to test and ensure compatibility with different hardware architectures (different memory capacities), we set the maximum input length to 128 and the maximum text generation length to 128. Furthermore, to meet the dimension requirements of \texttt{te.Linear}, we set the batch size to 8.

We use throughput as the evaluation metric, which represents the total text length that can be processed per second:
$Throughput = (input\_len + output\_len) / time$.

\begin{table}[!ht]
    \centering
    \caption{Inference Throughput (Tokens/s) for different model sizes on different GPUs and different data types}
    \label{tab:llm-inference}
    
    \begin{tabular}{ll rrr}
        \toprule
        \textbf{GPU} & \textbf{Model} & \textbf{FP32} & \textbf{BF16} & \textbf{FP8} \\
        \midrule
        
        \multirow{2}{*}{4090} & llama-3B & 653.33 & 638.19 & 641.19 \\
        & llama-2-7B & OOM & 526.65 & OOM \\
        \midrule 

        \multirow{3}{*}{A100} & llama-3B & 507.67 & 539.82 & - \\
        & llama-2-7B & 309.93 & 488.06 & - \\
        & llama-2-13B & OOM & 382.53 & - \\
        \midrule 

        \multirow{3}{*}{H800} & llama-3B & 546.55 & 506.28 & 505.74 \\
        & llama-2-7B & 359.69 & 432.91 & 428.15 \\
        & llama-2-13B & 214.06 & 344.25 & 343.25 \\
        
        \bottomrule
    \end{tabular}
\end{table}

\noindent\textbf{LLM inference throughput results.}
        We test the state-of-the-art decode-only models on inference with different data types, as shown in Table \ref{tab:llm-inference}. We set the input length and text generation length to be relatively short, and the decode-only model is memory-bound during inference, so the computational advantages of FP8 Tensor Cores are not significant. Moreover, since the current Transformer Engine does not provide comprehensive support, data transmission between modules still occurs in FP16/FP32 without operator fusion. It is possible that when the model size and input data length increase, and with good operator fusion support, a certain improvement can be achieved.
    


\begin{tcolorbox}[colback=green!5!white, colframe=green!50!black, title=Insight]
1. \texttt{wgmma} can achieve 95\% of Hopper's theoretical peak performance, while the backward-compatible \texttt{mma} can only reach 62.9\%.

2. Using \texttt{wgmma} requires attention to power wall constraints, whereas \texttt{mma} demonstrates better energy efficiency.

3. Utilizing 4th-generation Tensor Core low-precision formats (FP8) is beneficial for libraries and applications, but careful consideration must be given to problem size, precision conversion overhead and accuracy loss.

\end{tcolorbox}

\section{Distributed Shared Memory}
\label{sec:dsm}

The Hopper architecture features a direct SM-to-SM communication network within clusters, enabling threads from one thread block to access shared memory of another block, known as distributed shared memory (DSM). Additionally, for cases where shared memory demand restricts active block numbers on an SM, DSM can partition data within the same cluster, alleviating shared memory demand per block.
The programmability of DSM is facilitated through the CUDA C function, \texttt{cluster.map\_shared\_rank(SMEM, DST\_BLOCK\_RANK)}, returning the shared memory address of the target block. Here, \texttt{SMEM} represents the shared memory pointer, and \texttt{DST\_BLOCK\_RANK} is the target block rank in the cluster. This is compiled into PTX code \texttt{mapa}, which maps the address of the shared variable in the target block.
We benchmark DSM using three aspects: DSM latency, throughput under various scheduling strategies and access patterns, and histogram applications. This multi-faceted testing approach helps uncover the programming performance of DSM.

\subsection{Latency}
To measure inter-SM data transfer latency, we launch multiple blocks, each with one thread. We utilize \texttt{mov.u32 \%0, \%\%smid} PTX code to record the SM ID of each block, ensuring they run on different SMs. Our testing method is consistent with the uniform stride p-chase described in Section \ref{sec:mem_lat}.

As shown in Table \ref{tab:mem_lat}, when we directly access the shared memory, the latency is 29 cycles. However, when we access the block's local shared memory by the distributed shared memory interface, the latency reaches 33 cycles. The results indicate that even if we access the local shared memory and it is not required to leverage SM-to-SM network, the distributed shared memory still introduces some overhead. When the cluster size increases to two, the access latency between SMs is 181 cycles, a 30\% reduction compared to L2 cache. Notice that transferring data from one SM to another via global memory incurs a cost of 1110 cycles (one store and one load), utilizing DSM can reduce this latency by 6.13$\times$, which is close to the 7$\times$ reduction reported in NVIDIA's official documentation. 
When the cluster size increases from two to sixteen (the maximum cluster size in Hopper), the access latencies between different blocks range from 184 to 213 cycles.

\subsection{Throughput}
In the DSM throughput test, we benchmark the performance for both intra-SM and inter-SM scenarios separately.

For the intra-SM case, the theoretical throughput of shared memory per SM is calculated as $1755\,\text{MHz} \times 128\,\text{Bytes} = 225\,\text{GB/s}$. 
Using the DSM interface to access the block's own memory only reaches 205 GB/s, 80\% of peak performance. It is noteworthy that intra-SM shared memory access can also be performed without using the DSM interface, achieving over 99.8\% of theoretical performance, as illustrated in Table \ref{tab:mem_bw}. This indicates that there exists an additional overhead of accessing the block's shared memory through DSM.

\begin{figure}[htbp]
    \centering
    \includegraphics[width=0.8\linewidth]{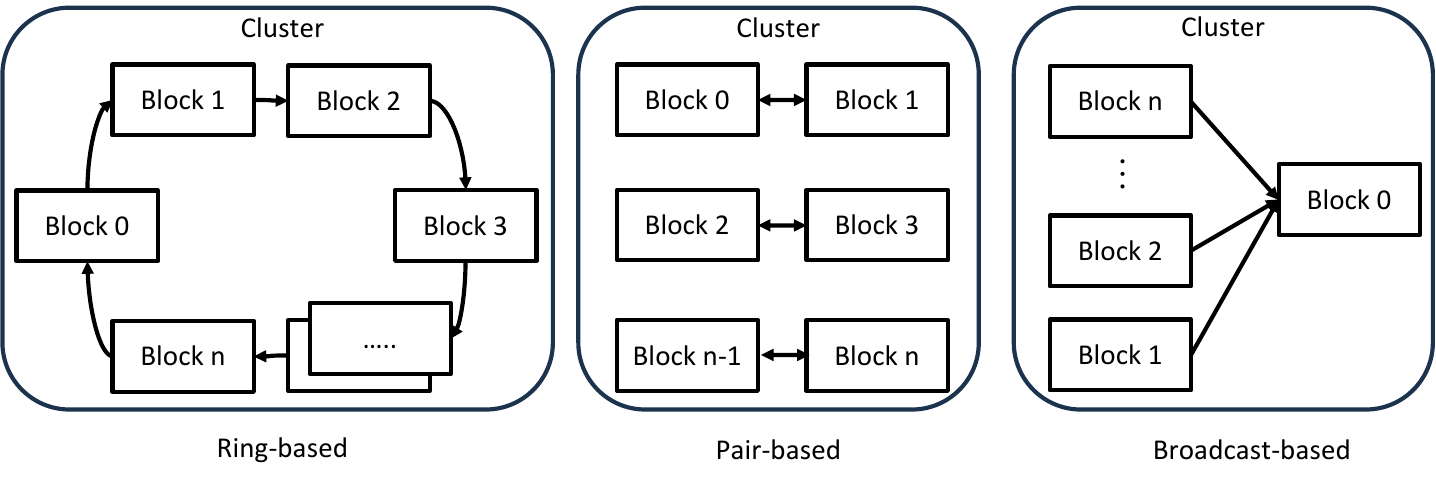}  
    \caption{Scenarios for DSM throughput testing}
    \label{fig:dsm_access_scen}
\end{figure}

For the inter-SM case, since DSM allows any block within a cluster to access another block, the access pattern significantly affects data movement throughput. Thus, as shown in Fig. \ref{fig:dsm_access_scen}, we designed three different copy scenarios to evaluate DSM throughput. The first is a ring-based access scheme, which is common in high-performance computing (e.g., ring allreduce \cite{thakur2005optimization,sergeev2018horovod}, matrix multiplication \cite{cannon1969cellular, chtchelkanova1997parallel}). In each cluster, threads in a block ranked by $R$ access the shared memory of the block ranked by $(R+1)\%CS$, where $CS$ is the cluster size. The second is pair-based, often used in butterfly communication scenarios in high-performance computing. The third is broadcast-based, a common access pattern in various parallel algorithms. In addition to above access pattern, we are also trying to explore the block scheduling policy within the cluster. There are three scheduling policies available in Hopper: default, Spread, and LoadBalancing. We explore the throughput of different access patterns under each scheduling policy.
In the remaining experiments in this section, we launch 50160 blocks to ensure sufficient block residency on the SMs. Additionally, we allocate 16KB of shared memory for each block.

\begin{figure}[htbp]
    \centering
    \includegraphics[width=0.8\linewidth]{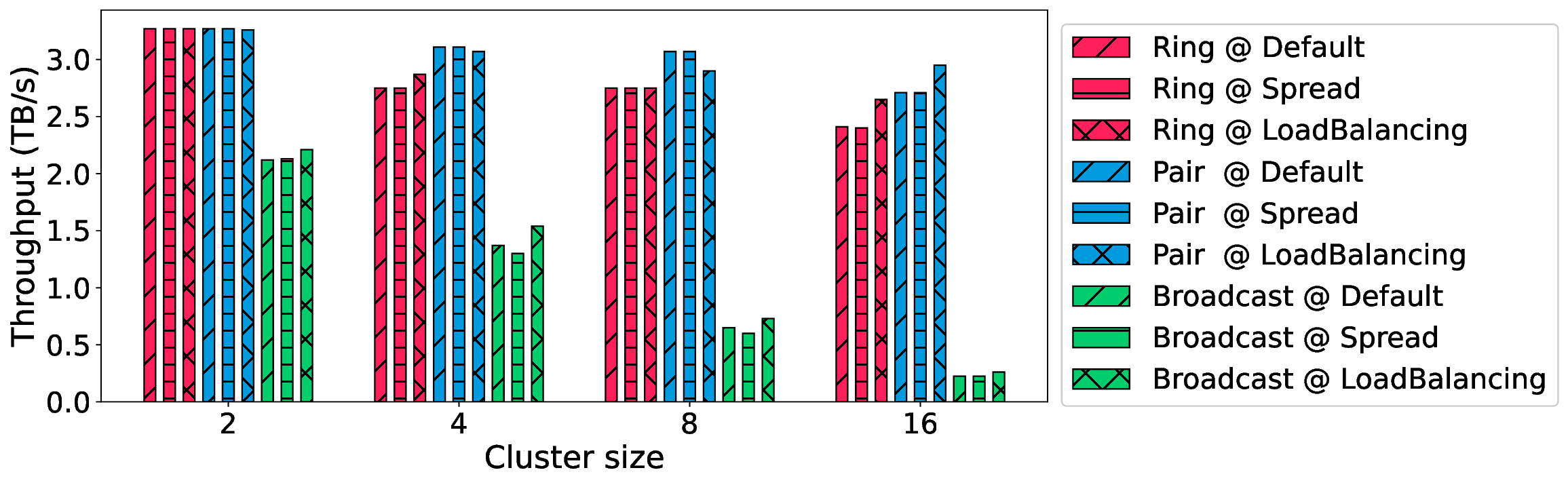}  
    \caption{Performance of access pattern on different policies}
    \label{fig:dsm_access_pattern}
\end{figure}

We observe the SM ID allocation for blocks within the same cluster. Under the Default and Spread scheduling policies, blocks from the same cluster are assigned to different SMs within a GPC. However, with LoadBalancing, blocks can be allocated to the same SM. As shown in Fig. \ref{fig:dsm_access_pattern}, LoadBalancing slightly outperforms the other two policies in most of cases. This may be because blocks on the same SM can access shared memory without using the SM-to-SM network, enhancing performance. Different access patterns exhibit significant throughput variations as cluster size increases. Ring-based and pair-based patterns show minimal differences, while Broadcast-based throughput decreases significantly with larger cluster sizes. In throughput tests, DSM does not achieve the original shared memory bandwidth.

\begin{figure}[htbp]
    \centering
    \includegraphics[width=0.7\linewidth]{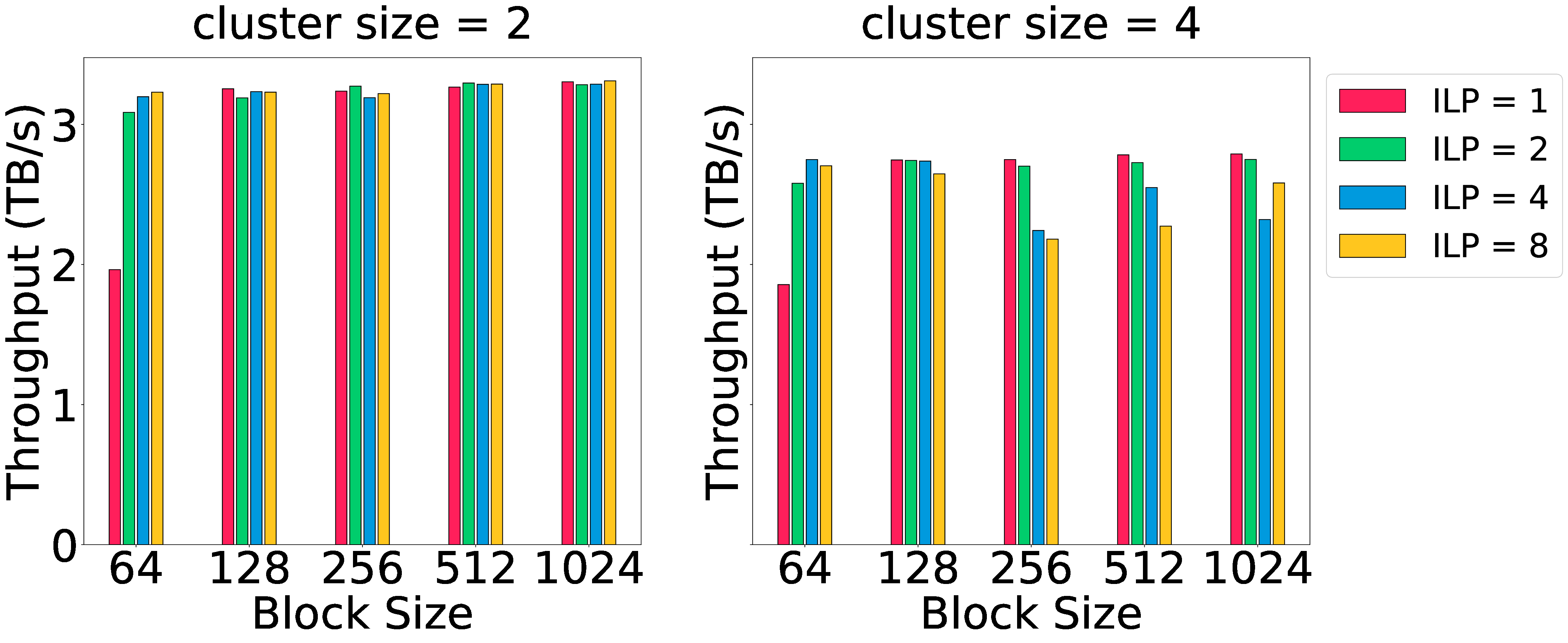}  
    \caption{The data communication throughput of the SM-to-SM network. ``ILP'' refers to the number of parallelizable data movement instructions.}
    \label{fig:dsm_throughput}
\end{figure}

We explore the impact of various parameters on DSM throughput by tuning cluster size, block size, and ILP. These parameters can guide task division in programming. We used the ring-based access pattern to evaluate the impact of different parameters. As shown in Fig. \ref{fig:dsm_throughput}, SM-to-SM throughput is illustrated for varying cluster and block sizes. As typically observed in similar benchmarks, when the block size is too small (e.g., 64), DSM performance is not fully utilized. However, increasing ILP can enhance SM-to-SM network utilization. When the block size is sufficiently large, ILP does not improve DSM performance. A peak throughput of nearly 3.28 TB/s is observed with a cluster size of 2, reducing to 2.78 TB/s with a cluster size of 4. Interestingly, as more blocks in the cluster compete for SM-to-SM bandwidth, the overall throughput gets lower and lower. While a larger cluster size can share shared memory for more blocks, it intensifies throughput competition. Balancing this tradeoff by selecting optimal block and cluster sizes is an important direction for exploration.

\subsection{Application-level}
We select histogram as an application-level benchmark.
Histograms are frequently employed in image processing and data mining to analyze the distribution of data elements by displaying the occurrence frequency of each element's value. While GPUs can accelerate histogram computation, achieving high computational efficiency is challenging.
We redesign the histogram\footnote{\url{https://github.com/NVIDIA/cuda-samples/Samples/2_Concepts_and_Techniques/histogram}} application using DSM, distributing bins across blocks in the same cluster. During histogram counting via shared memory, each thread loads the element and determines the DSM address for the target bin, followed by an atomic increment operation. The histogram application has the following two characteristics: 1) The data access flow is many-to-one similar to broadcasting (shared memory in each block is accessed by multiple blocks). 2) Due to the randomness of input data, writes to shared memory may generate unpredictable bank conflicts, which keeps ILP at a relatively low level. We adjust cluster size, block size, and bin count, measuring element processing throughput.

\begin{figure}[htbp]
    \centering
    \includegraphics[width=0.7\linewidth]{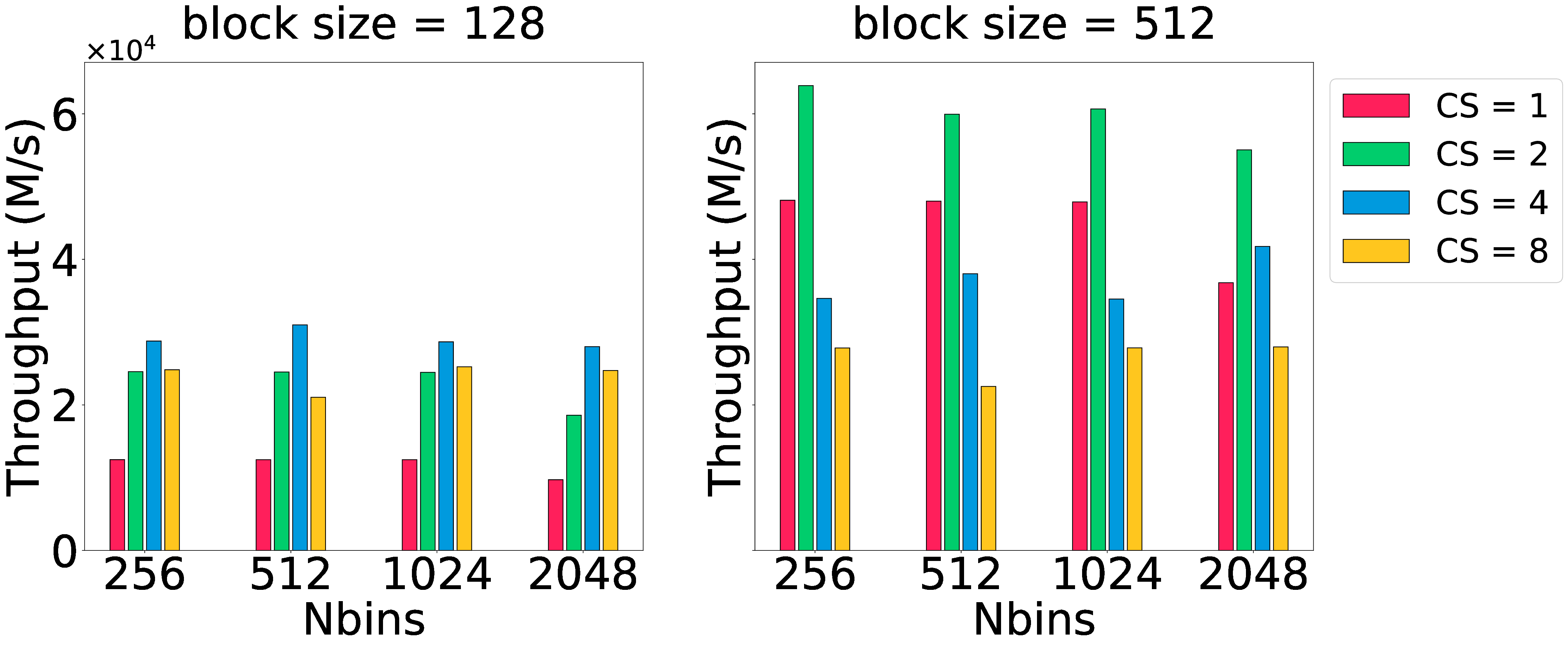}  
    \caption{Performance of the histogram application with distributed shared memory. The throughput is measured by the number of processing elements per second. ``CS'' refers to the cluster size. ``Nbins'' refers to the number of histogram bins.}
    \label{fig:histogram}
\end{figure}

The results shown in Fig. \ref{fig:histogram} can be summarized in two points. First, in addition to cluster size from 1 to 2, as cluster size increases, performance actually decreases.  This reflects the broadcast-like access pattern degrading performance with larger clusters, corresponding to the results in Fig. \ref{fig:dsm_access_pattern}. Second, increasing block size can significantly improve throughput. This demonstrates how higher block sizes can compensate for low ILP scenarios, which corresponds to the results in Fig. \ref{fig:dsm_throughput}. These findings demonstrate how the fundamental DSM behavior patterns we identified can be used to predict and understand real application performance, providing practical guidance for optimizing DSM-based applications.

Beyond the instruction-to-application analysis, we observe that the optimal cluster size differs for various block sizes (CS=4 for block size 128, CS=2 for block size 512). This suggests that kernel configuration parameters should be carefully selected to maximize performance. Additionally, when using DSM, we achieve over 30\% performance improvement compared to not using it (CS=1 for block size 512). Overall, choosing an appropriate cluster size eases the on-chip shared memory traffic by leveraging the SM-to-SM network resource, ultimately improving overall performance.



\begin{tcolorbox}[colback=green!5!white, colframe=green!50!black, title=Insight]
1. DSM reduces inter-SM communication latency by 6.1×.

2. DSM performance is highly dependent on access patterns. For broadcast-style access patterns, contention on shared memory can drastically degrade performance.

3. Kernel configuration parameters (e.g., cluster size and block size) significantly impact DSM performance, and should be carefully selected in practical applications. 

\end{tcolorbox}
\section{Dynamic Programming Instructions}
\label{sec:dpx}
NVIDIA offers DPX functions\footnote{\url{https://docs.nvidia.com/cuda/cuda-c-programming-guide/index.html##dpx}} from CUDA 12 onward to accelerate dynamic programming code, enhancing programming ease. On the latest Hopper architecture, these functions are hardware-accelerated. In this section, we conduct both instruction-level and application-level benchmarks to comprehensively evaluate the potential and real-world performance of DPX.

\subsection{Instruction-level}
Our instruction-level test of DPX functions focuses on the instruction latency and throughput.
For latency assessment, we utilize a thread to iteratively issue DPX functions, calculating their average latency. In the throughput test, we employ a block to repeatedly issue DPX functions, determining the DPX instruction throughput for each SM.
To pinpoint the location of DPX acceleration hardware, we vary the number of launched blocks and observed the relationship between DPX throughput and the launched block count.

\begin{figure*}[htb]
    \centering
    \includegraphics[width=0.98\linewidth]{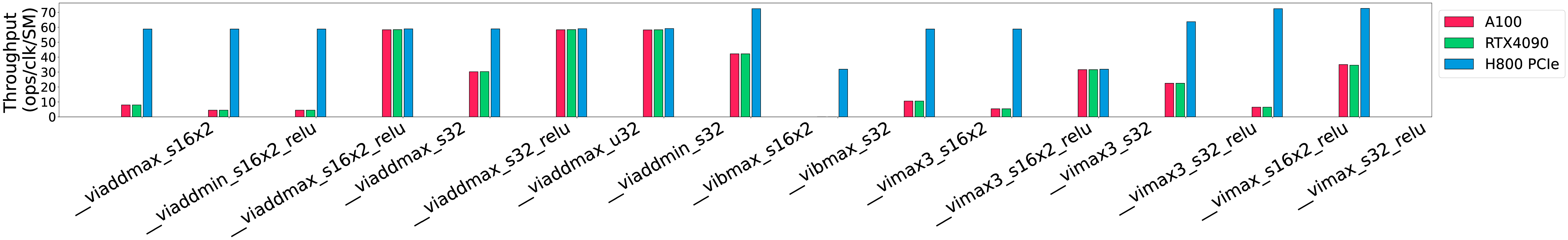} 
    \vspace{-1.2 em}
    \caption{Throughput of DPX functions on different devices}
    \label{fig:dpx_bw}
\end{figure*}

Fig. \ref{fig:dpx_bw} show the throughput of the DPX functions on three tested GPUs. The latency data is provided in the supplementary materials, which are available online.
Since the DPX of RTX4090 and A100 are software emulation, their performance is almost the same. What can be observed is that for \texttt{relu} instructions, the performance of H800 is significantly better than the other two. For 16-bit operations, H800 also has significant acceleration, up to 13 times.

However, not all functions have acceleration effects on Hopper. For some simple operations (e.g. \texttt{\_\_viaddmax\_\texttt{S32}}, which accepts 3 signed integers \texttt{(s1, s2, s3)} and returns \texttt{max(s1+s2,s3)}), we find that the performance of the three devices is close. In fact, by observing the SASS code, we find that new instructions (\texttt{VIMNMX}) are used on Hopper. Compared with previous \texttt{IMNMX}, performance does not seem to improve significantly. But in general, the Hopper architecture with DPX hardware acceleration has better performance than the previous generation architecture.

Additionally, \texttt{\_\_vibmax\_\texttt{S32}} data is not available on RTX4090 and A100. The reason is that compilation optimization optimizes this function into a max instruction. If we want to prevent this optimization, throughput measurements will be greatly affected. 

Another finding is that on H800, when the number of launched blocks is less than the number of SMs, the throughput of DPX functions is proportional to the number of blocks. When the number of blocks just exceeds an integral multiple of the number of SMs, the throughput plummets, gradually returning to the maximum level as the number of blocks increases. Maximum throughput occurs when the number of blocks is an integer multiple of the number of SMs. Therefore, we have enough reason to infer that the DPX acceleration unit is located at the SM level.


\subsection{Application-level}

This case study explores the application of DPX instructions to the Smith-Waterman algorithm\cite{smith1981identification}, a classic dynamic programming algorithm used in bioinformatics. The Smith-Waterman algorithm finds optimal local alignments between two sequences and has wide-ranging applications in sequence alignment, genome analysis, and protein structure prediction. 
It employs dynamic programming by filling a two-dimensional matrix to compute the optimal local alignment score. Each matrix element represents the optimal local alignment score at corresponding positions within the two sequences. The matrix is filled recursively using a specific formula, ultimately yielding the optimal alignment.

The Smith-Waterman algorithm finds optimal local alignments between sequences $Q = (q_0 q_1 \dots q_{m-1})$ and $S = (s_0 s_1 \dots s_{n-1})$ using dynamic programming. The core recurrence relation, calculating the score $H(i,j)$, is:
$$H(i,j) = \max
\begin{cases}
H(i-1, j-1) + \sigma(q_{i-1}, s_{j-1}) \\
E(i,j) \\
F(i,j) \\
0
\end{cases}$$
where $\sigma$ is the substitution scoring function. Affine gap penalties are incorporated via $E(i,j)$ and $F(i,j)$:
$$E(i,j) = \max
\begin{cases}
E(i-1,j) - \beta \\
H(i-1,j) - \alpha
\end{cases}$$
$$F(i,j) = \max
\begin{cases}
F(i,j-1) - \beta \\
H(i,j-1) - \alpha
\end{cases}$$
with $\alpha$ as the gap opening penalty and $\beta$ as the gap extension penalty. The matrix $H$ is initialized with zeros along the first row and column. The optimal local alignment score is the maximum value in $H$. This computation requires $O(mn)$ time and $O(\min(m,n))$ space.

We can see that the computation of $H(i, j)$ is well-suited for the \texttt{\_\_vimax3\_datatype\_relu} instruction. This instruction takes three numbers \texttt{(s1, s2, s3)} as input and returns \texttt{max(s1, s2, s3, 0)}. The calculation of $E(i, j)$ and $F(i, j)$ can benefit from the \texttt{\_\_viaddmax\_datatype} instruction, which takes three numbers \texttt{(s1, s2, s3)} and returns \texttt{max(s1 + s2, s3)}. The Smith-Waterman algorithm primarily involves repeatedly calculating $H$, $E$, and $F$. Therefore, these DPX instructions can accelerate these key operations, making them highly suitable for optimizing the algorithm’s performance.

Our testing methodology is based on \cite{schmidt2024cudasw++}. We use the same code, data, and testing parameter configuration as described in \cite{schmidt2024cudasw++}. We use 20 protein sequences, with lengths ranging from 144 to 5478, as queries against a simulated databases. The database contains one million sequences, with fixed lengths 1024. Our results represent the average performance across all twenty queries. The performance metric for our experiments is GCUPS (Giga Cell Updates Per Second), representing the number of dynamic programming matrix cell updates performed per second. 
\begin{figure}[htbp]
    \centering
    \includegraphics[width=0.6\linewidth]{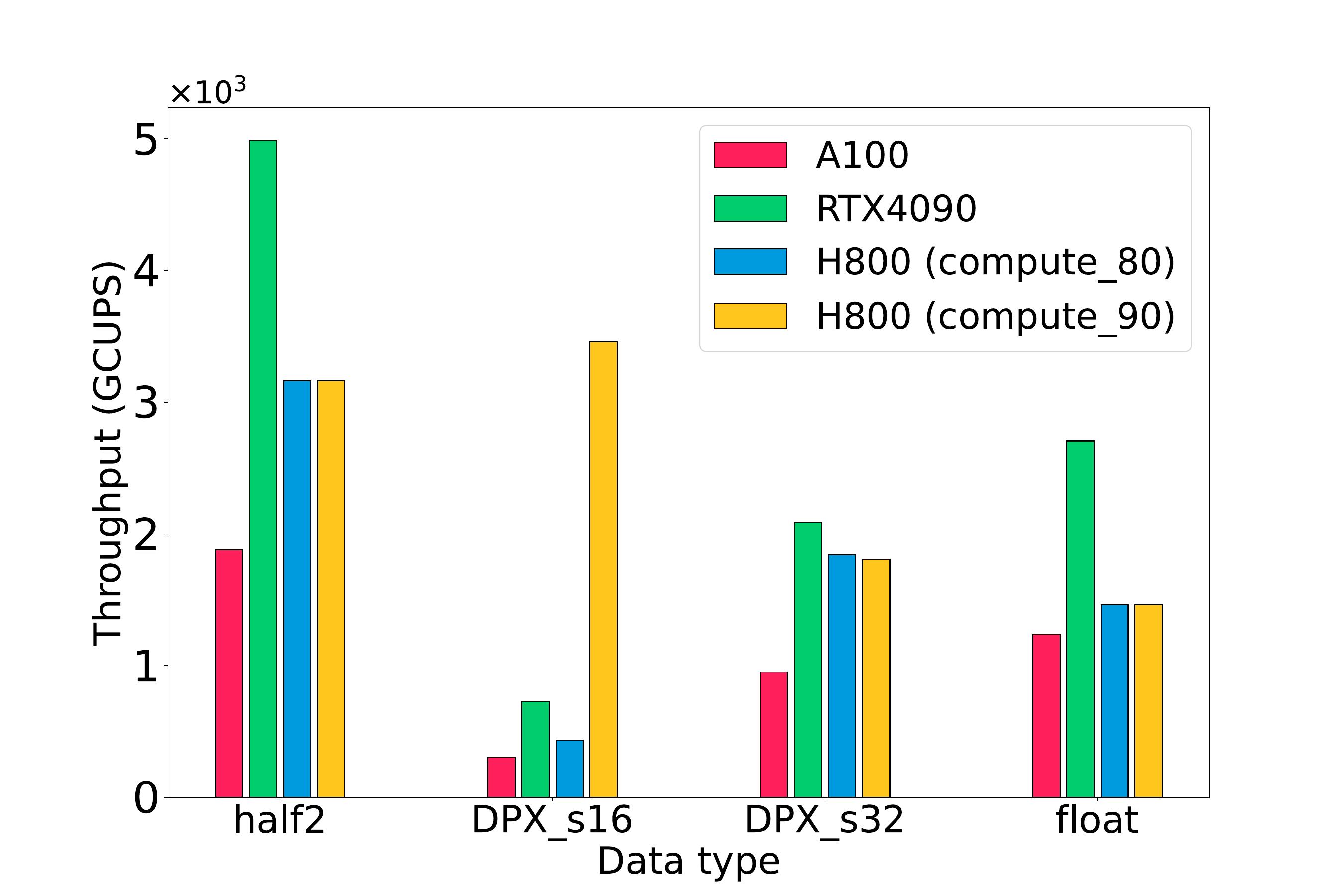}
    \caption{Performance of the Smith-Waterman algorithm across various data types}
    \label{fig:dpx_app}
\end{figure}

To further investigate the hardware acceleration benefits of DPX, we compare performance not only across different devices but also on the H800 running under both Ampere and Hopper virtual architectures. This comparison highlights Hopper's dedicated acceleration support for DPX instructions. In addition to the 16-bit and 32-bit integer types supported by DPX, we also considered half-precision (\texttt{half2}) and single-precision (\texttt{float}) floating-point types. As shown in Figure \ref{fig:dpx_app}, best performance is achieved on the RTX 4090 using half-precision data, reaching 4987.4 GCUPS. This superior performance is attributable to the RTX 4090's higher core clock speeds and greater number of SMs. A similar trend is observed for both \texttt{half2} and \texttt{float} data types, neither of which leverage DPX instructions. 

For the data types supported by DPX, the performance with signed 32-bit integers (\texttt{S32}) follows a similar pattern: RTX 4090 \textgreater{} H800 \textgreater{} A100. This \texttt{S32} performance aligns with our instruction-level benchmark results, indicating a limited or even no performance gain from DPX for some 32-bit instructions. Furthermore, compiling the DPX function for \texttt{compute\_80} (without hardware acceleration) versus \texttt{compute\_90} (with hardware acceleration) shows no significant difference in performance for \texttt{S32}. However, for signed 16-bit integers (\texttt{S16}), the code generated with \texttt{compute\_90} performs significantly better, benefiting from Hopper's hardware acceleration support for DPX instructions. This advantage over compute \texttt{compute\_80}, A100, and RTX4090, is consistent with our instruction-level benchmarks, which demonstrate a significant performance improvement for 16-bit integer instructions on the Hopper architecture.

In summary, this application-level testing provides several insights into performance optimization for dynamic programming applications:
1) Using lower-precision data types, whether integer or floating-point, offers significant performance advantages on the Hopper architecture. Where the reduced precision satisfies the application's data representation requirements, its use should be prioritized.
2) For architectures prior to Hopper, avoid using \texttt{S16} as a data type unless memory constraints dictate otherwise. Due to the greater availability of hardware units for processing floating-point, prioritize floating-point data types when possible.
3) When considering DPX instructions, the results from instruction-level benchmarks can guide application performance optimization. Not all DPX instructions offer performance advantages on the Hopper architecture.

\begin{tcolorbox}[colback=green!5!white, colframe=green!50!black, title=Insight]
1. More complex operations (e.g., those with more operands and ReLU operations) benefit more significantly from DPX instructions.

2. The acceleration effect of 16-bit DPX instructions is more obvious, while some 32-bit operations have no acceleration effect.

\end{tcolorbox}
\section{Conclusion}

This study presents a comprehensive evaluation of NVIDIA's Hopper GPU architecture through multi-level benchmarking, revealing significant performance improvements and optimization opportunities. Our results demonstrate that Hopper's memory subsystem achieves 2.6× higher L2 cache throughput compared to Ampere, while its innovative Distributed Shared Memory (DSM) reduces inter-SM communication latency by 6.1×. These enhancements are critical for memory-bound workloads in AI and high-performance computing.

The fourth-generation tensor cores show substantial gains when properly utilized. While backward-compatible \texttt{mma} instructions achieve only 62.9\% of peak performance on average, the dedicated \texttt{wgmma} instructions fully unlock Hopper's computational potential. Furthermore, the FP8 data type delivers 2× higher throughput than FP16 for large-scale operations, though its benefits diminish for smaller workloads due to conversion overhead.

Specialized hardware units like the TMA and DPX instructions further enhance performance. TMA enables efficient asynchronous execution, accelerating matrix multiplication by 50\%, while DPX instructions improve dynamic programming workloads by up to 13× in microbenchmarks. Real-world applications, such as the Smith-Waterman algorithm, benefit significantly, achieving a 4.75× speedup with 16-bit integers.

These findings provide actionable insights for developers. To maximize Hopper's performance, we recommend prioritizing \texttt{wgmma} over \texttt{mma} instructions, adopting FP8 for large-scale computations, leveraging DSM for efficient inter-SM communication, and utilizing DPX for dynamic programming tasks. Future work should explore optimal DSM configurations, precision-aware DPX optimizations, and the integration of TMA with other asynchronous execution features. Our research improves understanding of the latest architecture's characteristics and performance, aiding in optimized algorithm design and application performance.

\begin{acks}
This work was partially supported by the National Natural Science Foundation of China (No. 62302126), the Shenzhen Science and Technology Program (No. RCBS20221008093125065), Guangdong Provincial Key Laboratory of Novel Security Intelligence Technologies (2022B1212010005), the Hong Kong RIF grant (No. R6021-20), and the Hong Kong CRF grant (No. C2004-21G, No. C7004-22G).
\end{acks}

\bibliographystyle{ACM-Reference-Format}
\bibliography{sample-base}


\begin{thebibliography}{60}


\ifx \showCODEN    \undefined \def \showCODEN     #1{\unskip}     \fi
\ifx \showISBNx    \undefined \def \showISBNx     #1{\unskip}     \fi
\ifx \showISBNxiii \undefined \def \showISBNxiii  #1{\unskip}     \fi
\ifx \showISSN     \undefined \def \showISSN      #1{\unskip}     \fi
\ifx \showLCCN     \undefined \def \showLCCN      #1{\unskip}     \fi
\ifx \shownote     \undefined \def \shownote      #1{#1}          \fi
\ifx \showarticletitle \undefined \def \showarticletitle #1{#1}   \fi
\ifx \showURL      \undefined \def \showURL       {\relax}        \fi
\providecommand\bibfield[2]{#2}
\providecommand\bibinfo[2]{#2}
\providecommand\natexlab[1]{#1}
\providecommand\showeprint[2][]{arXiv:#2}

\bibitem[Abdelkhalik et~al\mbox{.}(2022)]%
        {abdelkhalik_hpec_2022}
\bibfield{author}{\bibinfo{person}{Hamdy Abdelkhalik}, \bibinfo{person}{Yehia Arafa}, \bibinfo{person}{Nandakishore Santhi}, {and} \bibinfo{person}{Abdel-Hameed~A. Badawy}.} \bibinfo{year}{2022}\natexlab{}.
\newblock \showarticletitle{Demystifying the {Nvidia Ampere} Architecture through Microbenchmarking and Instruction-level Analysis}. In \bibinfo{booktitle}{\emph{2022 IEEE High Performance Extreme Computing Conference (HPEC)}}. \bibinfo{pages}{1--8}.
\newblock


\bibitem[Arafa et~al\mbox{.}(2019)]%
        {arafa2019low}
\bibfield{author}{\bibinfo{person}{Yehia Arafa}, \bibinfo{person}{Abdel-Hameed~A Badawy}, \bibinfo{person}{Gopinath Chennupati}, \bibinfo{person}{Nandakishore Santhi}, {and} \bibinfo{person}{Stephan Eidenbenz}.} \bibinfo{year}{2019}\natexlab{}.
\newblock \showarticletitle{Low overhead instruction latency characterization for nvidia gpgpus}. In \bibinfo{booktitle}{\emph{2019 IEEE High Performance Extreme Computing Conference (HPEC)}}. IEEE, \bibinfo{pages}{1--8}.
\newblock


\bibitem[Arafa et~al\mbox{.}(2020)]%
        {arafa2020verified}
\bibfield{author}{\bibinfo{person}{Yehia Arafa}, \bibinfo{person}{Ammar ElWazir}, \bibinfo{person}{Abdelrahman ElKanishy}, \bibinfo{person}{Youssef Aly}, \bibinfo{person}{Ayatelrahman Elsayed}, \bibinfo{person}{Abdel-Hameed Badawy}, \bibinfo{person}{Gopinath Chennupati}, \bibinfo{person}{Stephan Eidenbenz}, {and} \bibinfo{person}{Nandakishore Santhi}.} \bibinfo{year}{2020}\natexlab{}.
\newblock \showarticletitle{Verified instruction-level energy consumption measurement for nvidia gpus}. In \bibinfo{booktitle}{\emph{Proceedings of the 17th ACM International Conference on Computing Frontiers}}. \bibinfo{pages}{60--70}.
\newblock


\bibitem[Bakhoda et~al\mbox{.}(2009)]%
        {bakhoda2009analyzing}
\bibfield{author}{\bibinfo{person}{Ali Bakhoda}, \bibinfo{person}{George~L Yuan}, \bibinfo{person}{Wilson~WL Fung}, \bibinfo{person}{Henry Wong}, {and} \bibinfo{person}{Tor~M Aamodt}.} \bibinfo{year}{2009}\natexlab{}.
\newblock \showarticletitle{Analyzing CUDA workloads using a detailed GPU simulator}. In \bibinfo{booktitle}{\emph{2009 IEEE international symposium on performance analysis of systems and software}}. IEEE, \bibinfo{pages}{163--174}.
\newblock


\bibitem[Braun et~al\mbox{.}(2020)]%
        {braun2020simple}
\bibfield{author}{\bibinfo{person}{Lorenz Braun}, \bibinfo{person}{Sotirios Nikas}, \bibinfo{person}{Chen Song}, \bibinfo{person}{Vincent Heuveline}, {and} \bibinfo{person}{Holger Fr{\"o}ning}.} \bibinfo{year}{2020}\natexlab{}.
\newblock \showarticletitle{A simple model for portable and fast prediction of execution time and power consumption of GPU kernels}.
\newblock \bibinfo{journal}{\emph{ACM Transactions on Architecture and Code Optimization (TACO)}} \bibinfo{volume}{18}, \bibinfo{number}{1} (\bibinfo{year}{2020}), \bibinfo{pages}{1--25}.
\newblock


\bibitem[Braun et~al\mbox{.}(2021)]%
        {taco2021_portable}
\bibfield{author}{\bibinfo{person}{Lorenz Braun}, \bibinfo{person}{Sotirios Nikas}, \bibinfo{person}{Chen Song}, \bibinfo{person}{Vincent Heuveline}, {and} \bibinfo{person}{Holger Fr\"{o}ning}.} \bibinfo{year}{2021}\natexlab{}.
\newblock \showarticletitle{A Simple Model for Portable and Fast Prediction of Execution Time and Power Consumption of {GPU} Kernels}.
\newblock \bibinfo{journal}{\emph{ACM Transactions Architecture and Code Optimization}} \bibinfo{volume}{18}, \bibinfo{number}{1}, Article \bibinfo{articleno}{7} (\bibinfo{date}{dec} \bibinfo{year}{2021}), \bibinfo{numpages}{25}~pages.
\newblock


\bibitem[Cannon(1969)]%
        {cannon1969cellular}
\bibfield{author}{\bibinfo{person}{Lynn~Elliot Cannon}.} \bibinfo{year}{1969}\natexlab{}.
\newblock \bibinfo{booktitle}{\emph{A cellular computer to implement the Kalman filter algorithm}}.
\newblock \bibinfo{publisher}{Montana State University}.
\newblock


\bibitem[Chtchelkanova et~al\mbox{.}(1997)]%
        {chtchelkanova1997parallel}
\bibfield{author}{\bibinfo{person}{Almadena Chtchelkanova}, \bibinfo{person}{John Gunnels}, \bibinfo{person}{Greg Morrow}, \bibinfo{person}{James Overfelt}, {and} \bibinfo{person}{Robert~A van~de Geijn}.} \bibinfo{year}{1997}\natexlab{}.
\newblock \showarticletitle{Parallel implementation of BLAS: General techniques for level 3 BLAS}.
\newblock \bibinfo{journal}{\emph{Concurrency: Practice and Experience}} \bibinfo{volume}{9}, \bibinfo{number}{9} (\bibinfo{year}{1997}), \bibinfo{pages}{837--857}.
\newblock


\bibitem[Corporation(2023)]%
        {nvidia-ptx-doc}
\bibfield{author}{\bibinfo{person}{NVIDIA Corporation}.} \bibinfo{year}{2023}\natexlab{}.
\newblock \bibinfo{booktitle}{\emph{{CUDA} Documentation: Parallel Thread Execution}}.
\newblock
\urldef\tempurl%
\url{https://docs.nvidia.com/cuda/parallel-thread-execution/index.html}
\showURL{%
\tempurl}


\bibitem[Dao et~al\mbox{.}(2022)]%
        {flashattention}
\bibfield{author}{\bibinfo{person}{Tri Dao}, \bibinfo{person}{Dan Fu}, \bibinfo{person}{Stefano Ermon}, \bibinfo{person}{Atri Rudra}, {and} \bibinfo{person}{Christopher R{\'e}}.} \bibinfo{year}{2022}\natexlab{}.
\newblock \showarticletitle{Flashattention: Fast and memory-efficient exact attention with {IO}-awareness}.
\newblock \bibinfo{journal}{\emph{Advances in Neural Information Processing Systems}}  \bibinfo{volume}{35} (\bibinfo{year}{2022}), \bibinfo{pages}{16344--16359}.
\newblock


\bibitem[Fan et~al\mbox{.}(2019)]%
        {fan2019predictable}
\bibfield{author}{\bibinfo{person}{Kaijie Fan}, \bibinfo{person}{Biagio Cosenza}, {and} \bibinfo{person}{Ben Juurlink}.} \bibinfo{year}{2019}\natexlab{}.
\newblock \showarticletitle{Predictable GPUs frequency scaling for energy and performance}. In \bibinfo{booktitle}{\emph{Proceedings of the 48th International Conference on Parallel Processing}}. \bibinfo{pages}{1--10}.
\newblock


\bibitem[Fasi et~al\mbox{.}(2021)]%
        {fasi2021numerical}
\bibfield{author}{\bibinfo{person}{Massimiliano Fasi}, \bibinfo{person}{Nicholas~J Higham}, \bibinfo{person}{Mantas Mikaitis}, {and} \bibinfo{person}{Srikara Pranesh}.} \bibinfo{year}{2021}\natexlab{}.
\newblock \showarticletitle{Numerical behavior of {NVIDIA} tensor cores}.
\newblock \bibinfo{journal}{\emph{PeerJ Computer Science}}  \bibinfo{volume}{7} (\bibinfo{year}{2021}), \bibinfo{pages}{e330}.
\newblock


\bibitem[Floridi and Chiriatti(2020)]%
        {floridi2020gpt}
\bibfield{author}{\bibinfo{person}{Luciano Floridi} {and} \bibinfo{person}{Massimo Chiriatti}.} \bibinfo{year}{2020}\natexlab{}.
\newblock \showarticletitle{GPT-3: Its nature, scope, limits, and consequences}.
\newblock \bibinfo{journal}{\emph{Minds and Machines}} \bibinfo{volume}{30}, \bibinfo{number}{4} (\bibinfo{year}{2020}), \bibinfo{pages}{681--694}.
\newblock


\bibitem[Guerreiro et~al\mbox{.}(2018)]%
        {guerreiro2018gpgpu}
\bibfield{author}{\bibinfo{person}{Joao Guerreiro}, \bibinfo{person}{Aleksandar Ilic}, \bibinfo{person}{Nuno Roma}, {and} \bibinfo{person}{Pedro Tomas}.} \bibinfo{year}{2018}\natexlab{}.
\newblock \showarticletitle{GPGPU power modeling for multi-domain voltage-frequency scaling}. In \bibinfo{booktitle}{\emph{2018 IEEE International Symposium on High Performance Computer Architecture (HPCA)}}. IEEE, \bibinfo{pages}{789--800}.
\newblock


\bibitem[Guerreiro et~al\mbox{.}(2019)]%
        {guerreiro2019modeling}
\bibfield{author}{\bibinfo{person}{Jo{\~a}o Guerreiro}, \bibinfo{person}{Aleksandar Ilic}, \bibinfo{person}{Nuno Roma}, {and} \bibinfo{person}{Pedro Tom{\'a}s}.} \bibinfo{year}{2019}\natexlab{}.
\newblock \showarticletitle{Modeling and decoupling the GPU power consumption for cross-domain DVFS}.
\newblock \bibinfo{journal}{\emph{IEEE Transactions on Parallel and Distributed Systems}} \bibinfo{volume}{30}, \bibinfo{number}{11} (\bibinfo{year}{2019}), \bibinfo{pages}{2494--2506}.
\newblock


\bibitem[Ho and Wong(2017)]%
        {ho2017exploiting}
\bibfield{author}{\bibinfo{person}{Nhut-Minh Ho} {and} \bibinfo{person}{Weng-Fai Wong}.} \bibinfo{year}{2017}\natexlab{}.
\newblock \showarticletitle{Exploiting half precision arithmetic in Nvidia GPUs}. In \bibinfo{booktitle}{\emph{2017 IEEE High Performance Extreme Computing Conference (HPEC)}}. IEEE, \bibinfo{pages}{1--7}.
\newblock


\bibitem[Hong and Kim(2009)]%
        {hong2009analytical}
\bibfield{author}{\bibinfo{person}{Sunpyo Hong} {and} \bibinfo{person}{Hyesoon Kim}.} \bibinfo{year}{2009}\natexlab{}.
\newblock \showarticletitle{An analytical model for a GPU architecture with memory-level and thread-level parallelism awareness}. In \bibinfo{booktitle}{\emph{Proceedings of the 36th annual international symposium on Computer architecture}}. \bibinfo{pages}{152--163}.
\newblock


\bibitem[Hong and Kim(2010)]%
        {isca2010_hong}
\bibfield{author}{\bibinfo{person}{Sunpyo Hong} {and} \bibinfo{person}{Hyesoon Kim}.} \bibinfo{year}{2010}\natexlab{}.
\newblock \showarticletitle{An integrated {GPU} power and performance model}. In \bibinfo{booktitle}{\emph{International Symposium on Computer Architecture (ISCA)}}.
\newblock


\bibitem[Jia et~al\mbox{.}(2019)]%
        {jia2019dissecting}
\bibfield{author}{\bibinfo{person}{Zhe Jia}, \bibinfo{person}{Marco Maggioni}, \bibinfo{person}{Jeffrey Smith}, {and} \bibinfo{person}{Daniele~Paolo Scarpazza}.} \bibinfo{year}{2019}\natexlab{}.
\newblock \showarticletitle{Dissecting the {NVidia Turing T4 GPU} via microbenchmarking}.
\newblock \bibinfo{journal}{\emph{arXiv preprint arXiv:1903.07486}} (\bibinfo{year}{2019}).
\newblock


\bibitem[Jia et~al\mbox{.}(2018)]%
        {jia2018dissecting}
\bibfield{author}{\bibinfo{person}{Zhe Jia}, \bibinfo{person}{Marco Maggioni}, \bibinfo{person}{Benjamin Staiger}, {and} \bibinfo{person}{Daniele~P Scarpazza}.} \bibinfo{year}{2018}\natexlab{}.
\newblock \showarticletitle{Dissecting the {NVIDIA Volta GPU} architecture via microbenchmarking}.
\newblock \bibinfo{journal}{\emph{arXiv preprint arXiv:1804.06826}} (\bibinfo{year}{2018}).
\newblock


\bibitem[Khairy et~al\mbox{.}(2020)]%
        {accelsim_isca2020}
\bibfield{author}{\bibinfo{person}{Mahmoud Khairy}, \bibinfo{person}{Zhesheng Shen}, \bibinfo{person}{Tor~M. Aamodt}, {and} \bibinfo{person}{Timothy~G. Rogers}.} \bibinfo{year}{2020}\natexlab{}.
\newblock \showarticletitle{Accel-Sim: An Extensible Simulation Framework for Validated {GPU} Modeling}. In \bibinfo{booktitle}{\emph{2020 ACM/IEEE 47th Annual International Symposium on Computer Architecture (ISCA)}}. \bibinfo{pages}{473--486}.
\newblock


\bibitem[Kwon et~al\mbox{.}(2023)]%
        {vllm}
\bibfield{author}{\bibinfo{person}{Woosuk Kwon}, \bibinfo{person}{Zhuohan Li}, \bibinfo{person}{Siyuan Zhuang}, \bibinfo{person}{Ying Sheng}, \bibinfo{person}{Lianmin Zheng}, \bibinfo{person}{Cody~Hao Yu}, \bibinfo{person}{Joseph~E. Gonzalez}, \bibinfo{person}{Hao Zhang}, {and} \bibinfo{person}{Ion Stoica}.} \bibinfo{year}{2023}\natexlab{}.
\newblock \showarticletitle{Efficient Memory Management for Large Language Model Serving with PagedAttention}. In \bibinfo{booktitle}{\emph{Proceedings of the ACM SIGOPS 29th Symposium on Operating Systems Principles}}.
\newblock


\bibitem[Leng et~al\mbox{.}(2013)]%
        {gpuwattch_isca2013}
\bibfield{author}{\bibinfo{person}{Jingwen Leng}, \bibinfo{person}{Tayler Hetherington}, \bibinfo{person}{Ahmed ElTantawy}, \bibinfo{person}{Syed Gilani}, \bibinfo{person}{Nam~Sung Kim}, \bibinfo{person}{Tor~M. Aamodt}, {and} \bibinfo{person}{Vijay~Janapa Reddi}.} \bibinfo{year}{2013}\natexlab{}.
\newblock \showarticletitle{{GPUWattch}: Enabling Energy Optimizations in {GPGPUs}}. In \bibinfo{booktitle}{\emph{Proceedings of the 40th Annual International Symposium on Computer Architecture}} (Tel-Aviv, Israel) \emph{(\bibinfo{series}{ISCA '13})}. \bibinfo{pages}{487–498}.
\newblock


\bibitem[Luitjens(2013)]%
        {Luitjens2013CUDAPro}
\bibfield{author}{\bibinfo{person}{Justin Luitjens}.} \bibinfo{year}{2013}\natexlab{}.
\newblock \bibinfo{title}{{CUDA Pro Tip}: Increase Performance with Vectorized Memory Access}.
\newblock \bibinfo{howpublished}{\url{https://developer.nvidia.com/blog/cuda-pro-tip-increase-performance-with-vectorized-memory-access/}}.
\newblock
\newblock
\shownote{NVIDIA Technical Blog}.


\bibitem[Luo et~al\mbox{.}(2024)]%
        {10579250}
\bibfield{author}{\bibinfo{person}{Weile Luo}, \bibinfo{person}{Ruibo Fan}, \bibinfo{person}{Zeyu Li}, \bibinfo{person}{Dayou Du}, \bibinfo{person}{Qiang Wang}, {and} \bibinfo{person}{Xiaowen Chu}.} \bibinfo{year}{2024}\natexlab{}.
\newblock \showarticletitle{Benchmarking and Dissecting the Nvidia Hopper GPU Architecture}. In \bibinfo{booktitle}{\emph{2024 IEEE International Parallel and Distributed Processing Symposium (IPDPS)}}. \bibinfo{pages}{656--667}.
\newblock
\href{https://doi.org/10.1109/IPDPS57955.2024.00064}{doi:\nolinkurl{10.1109/IPDPS57955.2024.00064}}


\bibitem[Markidis et~al\mbox{.}(2018)]%
        {markidis_tensor_ipdpsw2018}
\bibfield{author}{\bibinfo{person}{Stefano Markidis}, \bibinfo{person}{Steven Wei~Der Chien}, \bibinfo{person}{Erwin Laure}, \bibinfo{person}{Ivy~Bo Peng}, {and} \bibinfo{person}{Jeffrey~S. Vetter}.} \bibinfo{year}{2018}\natexlab{}.
\newblock \showarticletitle{{NVIDIA} Tensor Core Programmability, Performance \& Precision}. In \bibinfo{booktitle}{\emph{2018 IEEE International Parallel and Distributed Processing Symposium Workshops (IPDPSW)}}. \bibinfo{pages}{522--531}.
\newblock


\bibitem[Martineau et~al\mbox{.}(2019)]%
        {benchmark_Martineau_eupar2019}
\bibfield{author}{\bibinfo{person}{Matt Martineau}, \bibinfo{person}{Patrick Atkinson}, {and} \bibinfo{person}{Simon McIntosh-Smith}.} \bibinfo{year}{2019}\natexlab{}.
\newblock \showarticletitle{Benchmarking the {NVIDIA V100 GPU} and Tensor Cores}. In \bibinfo{booktitle}{\emph{Euro-Par 2018: Parallel Processing Workshops}}. \bibinfo{publisher}{Springer International Publishing}, \bibinfo{address}{Cham}, \bibinfo{pages}{444--455}.
\newblock


\bibitem[Maslej et~al\mbox{.}(2023)]%
        {maslej2023artificial}
\bibfield{author}{\bibinfo{person}{Nestor Maslej}, \bibinfo{person}{Loredana Fattorini}, \bibinfo{person}{Erik Brynjolfsson}, \bibinfo{person}{John Etchemendy}, \bibinfo{person}{Katrina Ligett}, \bibinfo{person}{Terah Lyons}, \bibinfo{person}{James Manyika}, \bibinfo{person}{Helen Ngo}, \bibinfo{person}{Juan~Carlos Niebles}, \bibinfo{person}{Vanessa Parli}, {et~al\mbox{.}}} \bibinfo{year}{2023}\natexlab{}.
\newblock \showarticletitle{Artificial intelligence index report 2023}.
\newblock \bibinfo{journal}{\emph{arXiv preprint arXiv:2310.03715}} (\bibinfo{year}{2023}).
\newblock


\bibitem[Mei and Chu(2017)]%
        {mei_tpds_2017}
\bibfield{author}{\bibinfo{person}{Xinxin Mei} {and} \bibinfo{person}{Xiaowen Chu}.} \bibinfo{year}{2017}\natexlab{}.
\newblock \showarticletitle{Dissecting {GPU} Memory Hierarchy Through Microbenchmarking}.
\newblock \bibinfo{journal}{\emph{IEEE Transactions on Parallel and Distributed Systems}} \bibinfo{volume}{28}, \bibinfo{number}{1} (\bibinfo{year}{2017}), \bibinfo{pages}{72--86}.
\newblock


\bibitem[Mei et~al\mbox{.}(2017)]%
        {mei2017survey}
\bibfield{author}{\bibinfo{person}{Xinxin Mei}, \bibinfo{person}{Qiang Wang}, {and} \bibinfo{person}{Xiaowen Chu}.} \bibinfo{year}{2017}\natexlab{}.
\newblock \showarticletitle{A survey and measurement study of GPU DVFS on energy conservation}.
\newblock \bibinfo{journal}{\emph{Digital Communications and Networks}} \bibinfo{volume}{3}, \bibinfo{number}{2} (\bibinfo{year}{2017}), \bibinfo{pages}{89--100}.
\newblock


\bibitem[Mei et~al\mbox{.}(2014)]%
        {mei2014benchmarking}
\bibfield{author}{\bibinfo{person}{Xinxin Mei}, \bibinfo{person}{Kaiyong Zhao}, \bibinfo{person}{Chengjian Liu}, {and} \bibinfo{person}{Xiaowen Chu}.} \bibinfo{year}{2014}\natexlab{}.
\newblock \showarticletitle{Benchmarking the memory hierarchy of modern GPUs}. In \bibinfo{booktitle}{\emph{Network and Parallel Computing: 11th IFIP WG 10.3 International Conference, NPC 2014, Ilan, Taiwan, September 18-20, 2014. Proceedings 11}}. Springer, \bibinfo{pages}{144--156}.
\newblock


\bibitem[Nabavinejad et~al\mbox{.}(2022)]%
        {nabavinejad2022coordinated}
\bibfield{author}{\bibinfo{person}{Seyed~Morteza Nabavinejad}, \bibinfo{person}{Sherief Reda}, {and} \bibinfo{person}{Masoumeh Ebrahimi}.} \bibinfo{year}{2022}\natexlab{}.
\newblock \showarticletitle{Coordinated batching and DVFS for DNN inference on GPU accelerators}.
\newblock \bibinfo{journal}{\emph{IEEE transactions on parallel and distributed systems}} \bibinfo{volume}{33}, \bibinfo{number}{10} (\bibinfo{year}{2022}), \bibinfo{pages}{2496--2508}.
\newblock


\bibitem[NVIDIA(2022)]%
        {te}
\bibfield{author}{\bibinfo{person}{NVIDIA}.} \bibinfo{year}{2022}\natexlab{}.
\newblock \bibinfo{booktitle}{\emph{TransformerEngine}}.
\newblock
\urldef\tempurl%
\url{https://github.com/NVIDIA/TransformerEngine}
\showURL{%
\tempurl}


\bibitem[OpenAI({[n.\,d.]})]%
        {chatgpt}
\bibfield{author}{\bibinfo{person}{OpenAI}.} \bibinfo{year}{[n.\,d.]}\natexlab{}.
\newblock \bibinfo{booktitle}{\emph{Introducing {ChatGPT}}}.
\newblock
\urldef\tempurl%
\url{https://openai.com/blog/chatgpt}
\showURL{%
\tempurl}


\bibitem[Paszke et~al\mbox{.}(2019)]%
        {pytorch}
\bibfield{author}{\bibinfo{person}{Adam Paszke}, \bibinfo{person}{Sam Gross}, \bibinfo{person}{Francisco Massa}, \bibinfo{person}{Adam Lerer}, \bibinfo{person}{James Bradbury}, \bibinfo{person}{Gregory Chanan}, \bibinfo{person}{Trevor Killeen}, \bibinfo{person}{Zeming Lin}, \bibinfo{person}{Natalia Gimelshein}, \bibinfo{person}{Luca Antiga}, {et~al\mbox{.}}} \bibinfo{year}{2019}\natexlab{}.
\newblock \showarticletitle{Pytorch: An imperative style, high-performance deep learning library}.
\newblock \bibinfo{journal}{\emph{Advances in neural information processing systems}}  \bibinfo{volume}{32} (\bibinfo{year}{2019}).
\newblock


\bibitem[Raihan et~al\mbox{.}(2019)]%
        {raihan_ispass2019}
\bibfield{author}{\bibinfo{person}{Md~Aamir Raihan}, \bibinfo{person}{Negar Goli}, {and} \bibinfo{person}{Tor~M. Aamodt}.} \bibinfo{year}{2019}\natexlab{}.
\newblock \showarticletitle{Modeling Deep Learning Accelerator Enabled {GPUs}}. In \bibinfo{booktitle}{\emph{2019 IEEE International Symposium on Performance Analysis of Systems and Software (ISPASS)}}. \bibinfo{pages}{79--92}.
\newblock


\bibitem[Saavedra and Smith(1995)]%
        {Saavedra_Smith_1995}
\bibfield{author}{\bibinfo{person}{R.H. Saavedra} {and} \bibinfo{person}{A.J. Smith}.} \bibinfo{year}{1995}\natexlab{}.
\newblock \showarticletitle{Measuring cache and {TLB} performance and their effect on benchmark runtimes}.
\newblock \bibinfo{journal}{\emph{IEEE Trans. Comput.}} \bibinfo{volume}{44}, \bibinfo{number}{10} (\bibinfo{date}{Jan} \bibinfo{year}{1995}), \bibinfo{pages}{1223–1235}.
\newblock
\href{https://doi.org/10.1109/12.467697}{doi:\nolinkurl{10.1109/12.467697}}


\bibitem[Saavedra-Barrera(1992)]%
        {Saavedra_Barrera_1992}
\bibfield{author}{\bibinfo{person}{RafaelHector Saavedra-Barrera}.} \bibinfo{year}{1992}\natexlab{}.
\newblock \showarticletitle{{CPU} performance evaluation and execution time prediction using narrow spectrum benchmarking}.
\newblock  (\bibinfo{date}{Jan} \bibinfo{year}{1992}).
\newblock


\bibitem[Schmidt et~al\mbox{.}(2024)]%
        {schmidt2024cudasw++}
\bibfield{author}{\bibinfo{person}{Bertil Schmidt}, \bibinfo{person}{Felix Kallenborn}, \bibinfo{person}{Alejandro Chacon}, {and} \bibinfo{person}{Christian Hundt}.} \bibinfo{year}{2024}\natexlab{}.
\newblock \showarticletitle{CUDASW++ 4.0: ultra-fast GPU-based Smith--Waterman protein sequence database search}.
\newblock \bibinfo{journal}{\emph{BMC bioinformatics}} \bibinfo{volume}{25}, \bibinfo{number}{1} (\bibinfo{year}{2024}), \bibinfo{pages}{342}.
\newblock


\bibitem[Sergeev and Del~Balso(2018)]%
        {sergeev2018horovod}
\bibfield{author}{\bibinfo{person}{Alexander Sergeev} {and} \bibinfo{person}{Mike Del~Balso}.} \bibinfo{year}{2018}\natexlab{}.
\newblock \showarticletitle{Horovod: fast and easy distributed deep learning in TensorFlow}.
\newblock \bibinfo{journal}{\emph{arXiv preprint arXiv:1802.05799}} (\bibinfo{year}{2018}).
\newblock


\bibitem[Shazeer(2020)]%
        {swiglu}
\bibfield{author}{\bibinfo{person}{Noam Shazeer}.} \bibinfo{year}{2020}\natexlab{}.
\newblock \showarticletitle{{GLU} Variants Improve Transformer.}
\newblock \bibinfo{journal}{\emph{arXiv: Learning,arXiv: Learning}} (\bibinfo{date}{Feb} \bibinfo{year}{2020}).
\newblock


\bibitem[Shekofteh et~al\mbox{.}(2020)]%
        {8853389}
\bibfield{author}{\bibinfo{person}{S.-Kazem Shekofteh}, \bibinfo{person}{Hamid Noori}, \bibinfo{person}{Mahmoud Naghibzadeh}, \bibinfo{person}{Holger Fröning}, {and} \bibinfo{person}{Hadi~Sadoghi Yazdi}.} \bibinfo{year}{2020}\natexlab{}.
\newblock \showarticletitle{cCUDA: Effective Co-Scheduling of Concurrent Kernels on GPUs}.
\newblock \bibinfo{journal}{\emph{IEEE Transactions on Parallel and Distributed Systems}} \bibinfo{volume}{31}, \bibinfo{number}{4} (\bibinfo{year}{2020}), \bibinfo{pages}{766--778}.
\newblock
\href{https://doi.org/10.1109/TPDS.2019.2944602}{doi:\nolinkurl{10.1109/TPDS.2019.2944602}}


\bibitem[Smith et~al\mbox{.}(1981)]%
        {smith1981identification}
\bibfield{author}{\bibinfo{person}{Temple~F Smith}, \bibinfo{person}{Michael~S Waterman}, {et~al\mbox{.}}} \bibinfo{year}{1981}\natexlab{}.
\newblock \showarticletitle{Identification of common molecular subsequences}.
\newblock \bibinfo{journal}{\emph{Journal of molecular biology}} \bibinfo{volume}{147}, \bibinfo{number}{1} (\bibinfo{year}{1981}), \bibinfo{pages}{195--197}.
\newblock


\bibitem[Sun et~al\mbox{.}(2023)]%
        {sun_tpds_2023}
\bibfield{author}{\bibinfo{person}{Wei Sun}, \bibinfo{person}{Ang Li}, \bibinfo{person}{Tong Geng}, \bibinfo{person}{Sander Stuijk}, {and} \bibinfo{person}{Henk Corporaal}.} \bibinfo{year}{2023}\natexlab{}.
\newblock \showarticletitle{Dissecting Tensor Cores via Microbenchmarks: Latency, Throughput and Numeric Behaviors}.
\newblock \bibinfo{journal}{\emph{IEEE Transactions on Parallel and Distributed Systems}} \bibinfo{volume}{34}, \bibinfo{number}{1} (\bibinfo{year}{2023}), \bibinfo{pages}{246--261}.
\newblock


\bibitem[Svedin et~al\mbox{.}({[n.\,d.]})]%
        {heart_ampere_2021}
\bibfield{author}{\bibinfo{person}{Martin Svedin}, \bibinfo{person}{Steven W.~D. Chien}, \bibinfo{person}{Gibson Chikafa}, \bibinfo{person}{Niclas Jansson}, {and} \bibinfo{person}{Artur Podobas}.} \bibinfo{year}{[n.\,d.]}\natexlab{}.
\newblock \showarticletitle{Benchmarking the Nvidia GPU Lineage: From Early K80 to Modern A100 with Asynchronous Memory Transfers}. In \bibinfo{booktitle}{\emph{Proceedings of the 11th International Symposium on Highly Efficient Accelerators and Reconfigurable Technologies}} (Online, Germany) \emph{(\bibinfo{series}{HEART '21})}. Article \bibinfo{articleno}{9}, \bibinfo{numpages}{6}~pages.
\newblock


\bibitem[Tang et~al\mbox{.}(2019)]%
        {tang2019impact}
\bibfield{author}{\bibinfo{person}{Zhenheng Tang}, \bibinfo{person}{Yuxin Wang}, \bibinfo{person}{Qiang Wang}, {and} \bibinfo{person}{Xiaowen Chu}.} \bibinfo{year}{2019}\natexlab{}.
\newblock \showarticletitle{The impact of GPU DVFS on the energy and performance of deep learning: An empirical study}. In \bibinfo{booktitle}{\emph{Proceedings of the Tenth ACM International Conference on Future Energy Systems}}. \bibinfo{pages}{315--325}.
\newblock


\bibitem[te42kyfo({[n.\,d.]})]%
        {gpu_benches}
\bibfield{author}{\bibinfo{person}{te42kyfo}.} \bibinfo{year}{[n.\,d.]}\natexlab{}.
\newblock \bibinfo{title}{{GPU} benchmarks}.
\newblock \bibinfo{howpublished}{\url{https://github.com/te42kyfo/gpu-benches}}.
\newblock


\bibitem[Thakur et~al\mbox{.}(2005)]%
        {thakur2005optimization}
\bibfield{author}{\bibinfo{person}{Rajeev Thakur}, \bibinfo{person}{Rolf Rabenseifner}, {and} \bibinfo{person}{William Gropp}.} \bibinfo{year}{2005}\natexlab{}.
\newblock \showarticletitle{Optimization of collective communication operations in MPICH}.
\newblock \bibinfo{journal}{\emph{The International Journal of High Performance Computing Applications}} \bibinfo{volume}{19}, \bibinfo{number}{1} (\bibinfo{year}{2005}), \bibinfo{pages}{49--66}.
\newblock


\bibitem[Touvron et~al\mbox{.}(2023a)]%
        {llama1}
\bibfield{author}{\bibinfo{person}{Hugo Touvron}, \bibinfo{person}{Thibaut Lavril}, \bibinfo{person}{Gautier Izacard}, \bibinfo{person}{Xavier Martinet}, \bibinfo{person}{Marie-Anne Lachaux}, \bibinfo{person}{Timoth{\'e}e Lacroix}, \bibinfo{person}{Baptiste Rozi{\`e}re}, \bibinfo{person}{Naman Goyal}, \bibinfo{person}{Eric Hambro}, \bibinfo{person}{Faisal Azhar}, {et~al\mbox{.}}} \bibinfo{year}{2023}\natexlab{a}.
\newblock \showarticletitle{Llama: Open and efficient foundation language models}.
\newblock \bibinfo{journal}{\emph{arXiv preprint arXiv:2302.13971}} (\bibinfo{year}{2023}).
\newblock


\bibitem[Touvron et~al\mbox{.}(2023b)]%
        {llama2}
\bibfield{author}{\bibinfo{person}{Hugo Touvron}, \bibinfo{person}{Louis Martin}, \bibinfo{person}{Kevin Stone}, \bibinfo{person}{Peter Albert}, \bibinfo{person}{Amjad Almahairi}, \bibinfo{person}{Yasmine Babaei}, \bibinfo{person}{Nikolay Bashlykov}, \bibinfo{person}{Soumya Batra}, \bibinfo{person}{Prajjwal Bhargava}, \bibinfo{person}{Shruti Bhosale}, {et~al\mbox{.}}} \bibinfo{year}{2023}\natexlab{b}.
\newblock \showarticletitle{Llama 2: Open foundation and fine-tuned chat models}.
\newblock \bibinfo{journal}{\emph{arXiv preprint arXiv:2307.09288}} (\bibinfo{year}{2023}).
\newblock


\bibitem[van Stigt et~al\mbox{.}(2022)]%
        {van2022isolating}
\bibfield{author}{\bibinfo{person}{Rico van Stigt}, \bibinfo{person}{Stephen~Nicholas Swatman}, {and} \bibinfo{person}{Ana-Lucia Varbanescu}.} \bibinfo{year}{2022}\natexlab{}.
\newblock \showarticletitle{Isolating gpu architectural features using parallelism-aware microbenchmarks}. In \bibinfo{booktitle}{\emph{Proceedings of the 2022 ACM/SPEC on International Conference on Performance Engineering}}. \bibinfo{pages}{77--88}.
\newblock


\bibitem[Vaswani et~al\mbox{.}(2017)]%
        {vaswani2017attention}
\bibfield{author}{\bibinfo{person}{Ashish Vaswani}, \bibinfo{person}{Noam Shazeer}, \bibinfo{person}{Niki Parmar}, \bibinfo{person}{Jakob Uszkoreit}, \bibinfo{person}{Llion Jones}, \bibinfo{person}{Aidan~N Gomez}, \bibinfo{person}{{\L}ukasz Kaiser}, {and} \bibinfo{person}{Illia Polosukhin}.} \bibinfo{year}{2017}\natexlab{}.
\newblock \showarticletitle{Attention is all you need}.
\newblock \bibinfo{journal}{\emph{Advances in neural information processing systems}}  \bibinfo{volume}{30} (\bibinfo{year}{2017}).
\newblock


\bibitem[Wang and Chu(2020)]%
        {tpds2020_gpudvfs}
\bibfield{author}{\bibinfo{person}{Qiang Wang} {and} \bibinfo{person}{Xiaowen Chu}.} \bibinfo{year}{2020}\natexlab{}.
\newblock \showarticletitle{{GPGPU} Performance Estimation With Core and Memory Frequency Scaling}.
\newblock \bibinfo{journal}{\emph{IEEE Transactions on Parallel and Distributed Systems}} \bibinfo{volume}{31}, \bibinfo{number}{12} (\bibinfo{year}{2020}), \bibinfo{pages}{2865--2881}.
\newblock


\bibitem[Wang et~al\mbox{.}(2024)]%
        {wang2024dso}
\bibfield{author}{\bibinfo{person}{Qiang Wang}, \bibinfo{person}{Laiyi Li}, \bibinfo{person}{Weile Luo}, \bibinfo{person}{Yijia Zhang}, {and} \bibinfo{person}{Bingqiang Wang}.} \bibinfo{year}{2024}\natexlab{}.
\newblock \showarticletitle{DSO: A GPU Energy Efficiency Optimizer by Fusing Dynamic and Static Information}.
\newblock \bibinfo{journal}{\emph{arXiv preprint arXiv:2407.13096}} (\bibinfo{year}{2024}).
\newblock


\bibitem[Wang et~al\mbox{.}(2019)]%
        {hpca2019_hybrid}
\bibfield{author}{\bibinfo{person}{Xiebing Wang}, \bibinfo{person}{Kai Huang}, \bibinfo{person}{Alois Knoll}, {and} \bibinfo{person}{Xuehai Qian}.} \bibinfo{year}{2019}\natexlab{}.
\newblock \showarticletitle{A Hybrid Framework for Fast and Accurate {GPU} Performance Estimation through Source-Level Analysis and Trace-Based Simulation}. In \bibinfo{booktitle}{\emph{2019 IEEE International Symposium on High Performance Computer Architecture (HPCA)}}. \bibinfo{pages}{506--518}.
\newblock


\bibitem[Wang et~al\mbox{.}(2020)]%
        {wang2020benchmarking}
\bibfield{author}{\bibinfo{person}{Yuxin Wang}, \bibinfo{person}{Qiang Wang}, \bibinfo{person}{Shaohuai Shi}, \bibinfo{person}{Xin He}, \bibinfo{person}{Zhenheng Tang}, \bibinfo{person}{Kaiyong Zhao}, {and} \bibinfo{person}{Xiaowen Chu}.} \bibinfo{year}{2020}\natexlab{}.
\newblock \showarticletitle{Benchmarking the performance and energy efficiency of AI accelerators for AI training}. In \bibinfo{booktitle}{\emph{2020 20th IEEE/ACM International Symposium on Cluster, Cloud and Internet Computing (CCGRID)}}. IEEE, \bibinfo{pages}{744--751}.
\newblock


\bibitem[Wong et~al\mbox{.}(2010)]%
        {wong2010demystifying}
\bibfield{author}{\bibinfo{person}{Henry Wong}, \bibinfo{person}{Misel-Myrto Papadopoulou}, \bibinfo{person}{Maryam Sadooghi-Alvandi}, {and} \bibinfo{person}{Andreas Moshovos}.} \bibinfo{year}{2010}\natexlab{}.
\newblock \showarticletitle{Demystifying GPU microarchitecture through microbenchmarking}. In \bibinfo{booktitle}{\emph{2010 IEEE International Symposium on Performance Analysis of Systems \& Software (ISPASS)}}. IEEE, \bibinfo{pages}{235--246}.
\newblock


\bibitem[Yan et~al\mbox{.}(2020a)]%
        {yan_ipdps_2020}
\bibfield{author}{\bibinfo{person}{Da Yan}, \bibinfo{person}{Wei Wang}, {and} \bibinfo{person}{Xiaowen Chu}.} \bibinfo{year}{2020}\natexlab{a}.
\newblock \showarticletitle{Demystifying Tensor Cores to Optimize Half-Precision Matrix Multiply}. In \bibinfo{booktitle}{\emph{2020 IEEE International Parallel and Distributed Processing Symposium (IPDPS)}}. \bibinfo{pages}{634--643}.
\newblock


\bibitem[Yan et~al\mbox{.}(2020b)]%
        {yan_optimize_ppopp2020}
\bibfield{author}{\bibinfo{person}{Da Yan}, \bibinfo{person}{Wei Wang}, {and} \bibinfo{person}{Xiaowen Chu}.} \bibinfo{year}{2020}\natexlab{b}.
\newblock \showarticletitle{Optimizing Batched {Winograd} Convolution on {GPUs}}. In \bibinfo{booktitle}{\emph{Proceedings of the 25th ACM SIGPLAN Symposium on Principles and Practice of Parallel Programming}} (San Diego, California) \emph{(\bibinfo{series}{PPoPP '20})}. \bibinfo{pages}{32–44}.
\newblock


\bibitem[Zhang and Sennrich(2019)]%
        {rmsnorm}
\bibfield{author}{\bibinfo{person}{Biao Zhang} {and} \bibinfo{person}{Rico Sennrich}.} \bibinfo{year}{2019}\natexlab{}.
\newblock \showarticletitle{Root Mean Square Layer Normalization}.
\newblock \bibinfo{journal}{\emph{Neural Information Processing Systems,Neural Information Processing Systems}} (\bibinfo{date}{Dec} \bibinfo{year}{2019}).
\newblock
\href{https://doi.org/10.5167/uzh-177483}{doi:\nolinkurl{10.5167/uzh-177483}}


\end{thebibliography}

\appendix









\section{Supplementary data and analysis}\label{appendix2}

\subsection{DPX instructions}
\begin{figure*}[h]
    \centering
    \includegraphics[width=0.98\linewidth]{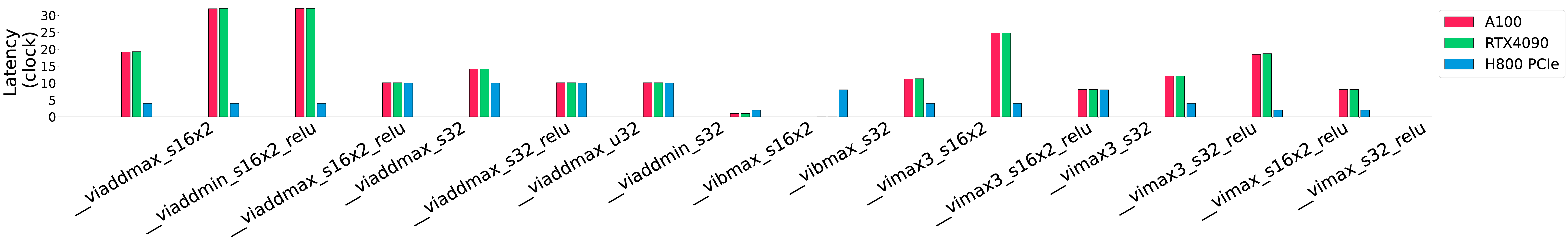} 
    \vspace{-1.2 em}
    \caption{Latency of DPX functions on different devices}
    \label{fig:dpx_lat}
    \vspace{-0.8 em}
\end{figure*}
Fig. \ref{fig:dpx_lat} demonstrates the performance advantages of Hopper's dedicated DPX hardware over software-emulated implementations. The H800 (Hopper) consistently shows the lowest latency across most DPX functions, while A100 and RTX4090 exhibit similar performance since both rely on software emulation. The most significant improvements occur in functions like \texttt{\_\_viaddmax\_s16x2} and certain vibmax operations, where Hopper achieves latency reductions of up to 70-80\%. However, the performance gains vary across different functions and data types, with 16-bit operations generally showing more pronounced benefits. This is consistent with the throughput results presented in our main text.

\subsection{Tensor Core}

\begin{figure*}[htbp]
    \centering
    \includegraphics[width=0.9\linewidth]{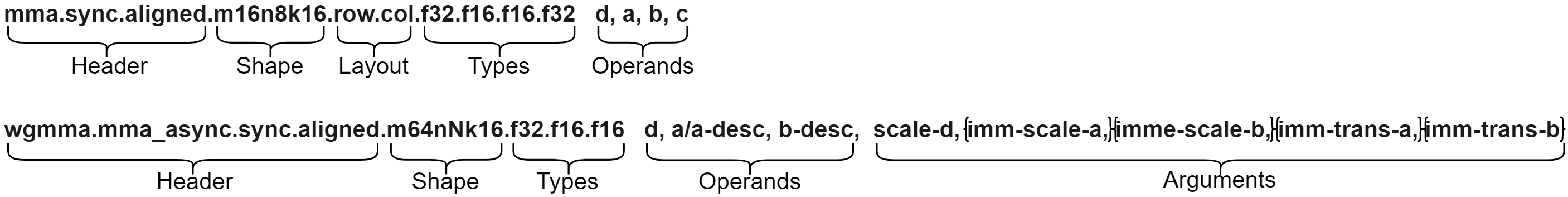}  
    \caption{The \texttt{mma} and \texttt{wgmma} instructions that perform $D = A \times B + C$ and $D = A \times B \{+ D\}$, respectively.}
    \label{fig:wgmma_inst}
\end{figure*}

\noindent\textbf{Difference between \texttt{wgmma} and \texttt{mma}.} Fig.~\ref{fig:wgmma_inst} provides examples of both \texttt{mma} and \texttt{wgmma} instructions, demonstrating mixed-precision capabilities.

The \texttt{wgmma} instructions implement warpgroup-level matrix multiply-and-accumulate operations through a structured six-step process shown in Listing \ref{list:wgmma}. First, matrices A, B, and D are loaded into registers or shared memory. Next, fence operations ensure memory consistency: \texttt{wgmma.fence} indicates data availability across the warpgroup. The core computation involves issuing asynchronous matrix operations using \texttt{wgmma.mma\_async} instructions, which execute within the async proxy framework. These operations are then grouped using \texttt{wgmma.commit\_group} to create a unified synchronization point. Finally, the system waits for wgmma-group completion, ensuring asynchronous matrix operations have finished executing before proceeding with subsequent computations. The transition from \texttt{mma} to \texttt{wgmma} represents a fundamental shift in programming paradigms. While \texttt{mma} instructions operate at warp-level (32 threads), \texttt{wgmma} extends operations to warpgroup-level (128 threads), requiring careful consideration of thread synchronization and data distribution patterns. This architectural evolution demands updated programming practices to fully exploit hardware capabilities.

\begin{lstlisting}[language=C++, caption={Programming Approach of \texttt{wgmma}},label={list:wgmma}, showstringspaces=false]
__global__ void wgmma_ptx(datatype* data, datatype* result) {

  // Load data into registers or into shared memory

  // asm inline start (ptx)
  wgmma.fence.sync.aligned;
  wgmma.mma_async...
  wgmma.commit_group.sync.aligned;
  wgmma.wait_group.sync.aligned remaining_group_num;
  // asm inline end (ptx)
}
\end{lstlisting}

\noindent\textbf{SASS analysis.}  We perform the disassembly of \texttt{mma} and \texttt{wgmma} instructions specifically for Hopper Tensor Cores, and the results are presented in Table~\ref{tab:sassanalysis}. The \texttt{mma} instructions undergo compilation into SASS instructions, with the naming convention following the established patterns: HMMA (for floating-point types), IMMA (for integer types), and BMMA (for binary types). 
Notably, there are two specialized types within \texttt{mma}: INT4 and FP8.

For INT4, on Ampere and Ada Tensor Cores, \texttt{mma} instructions are compiled into \texttt{IMMA.16832.S4.S4} instructions. However, a noteworthy deviation occurs on Hopper, where INT4 \texttt{mma} instructions are compiled into a series of IMAD instructions, which eventually run on the CUDA cores. 
This deviation results in performance that may fall short of the expected performance levels achievable with Tensor Cores. Additionally, while \texttt{mma} operations support FP32 accumulate when the data type of matrices A/B are FP8, \texttt{mma} do not support FP16 accumulate. The compiled SASS code remains \texttt{HMMA.16816.F32} in both the FP8/FP32 and FP16/FP32 cases.

The latest \texttt{wgmma} instructions are currently exclusive to Hopper Tensor Cores, despite NVIDIA's assertion that both Ada and Hopper feature fourth generation Tensor Cores. Unlike \texttt{mma}, \texttt{wgmma} instructions are compiled into the new \texttt{GMMA} SASS instructions. Users can program two variations of FP8 (E5M2 and E4M3) Tensor Cores using \texttt{wgmma}. 
Notably, as \texttt{wgmma} does not have forward compatibility requirements like \texttt{mma}, it does not support the INT4 data type.


\begin{table}[h]
    \centering
    \caption{SASS Instructions for Different Hopper Tensor Core PTX Instructions}
    \label{tab:sassanalysis}
    
    \begin{tabular}{llll}
        \toprule
        \textbf{A/B} & \textbf{C/D} & \textbf{mma} & \textbf{wgmma} \\
        \midrule
        
        FP16 & FP16 & \texttt{HMMA.16816.F16} & \texttt{HGMMA.64x256x16.F16} \\
        \midrule
        
        FP16 & FP32 & \texttt{HMMA.16816.F32} & \texttt{HGMMA.64x256x16.F32} \\
        \midrule
        
        TF32 & FP32 & \texttt{HMMA.1688.F32.TF32} & \texttt{HGMMA.64x256x8.F32.TF32} \\
        \midrule
        
        FP8 & FP16 & $\times$ & \makecell[l]{\texttt{QGMMA.64x256x32.F16.E5M2.E5M2} \\ \texttt{QGMMA.64x256x32.F16.E4M3.E4M3}} \\
        \midrule
        
        FP8 & FP32 & \texttt{HMMA.16816.F32} & \makecell[l]{\texttt{QGMMA.64x256x32.F32.E4M3.E4M3} \\ \texttt{QGMMA.64x256x32.F32.E5M2.E5M2}} \\
        \midrule
        
        INT8 & INT32 & \texttt{IMMA.16832.S8.S8} & \texttt{IGMMA.64x256x32.S8.S8} \\
        \midrule
        
        INT4 & INT32 & \texttt{IMAD.MOV.U32} & $\times$ \\
        \midrule
        
        Binary & INT32 & \texttt{BMMA.168256.AND.POPC} & \texttt{BGMMA.64x256x256.AND.POPC} \\
        
        \bottomrule
    \end{tabular}
\end{table}

\noindent\textbf{Sparse \texttt{wgmma} Instructions Performance.} Table \ref{tab:sparsewgmma} presents the detailed performance characteristics of sparse wgmma instructions on H800 Tensor Cores, complementing the dense wgmma results shown in main text. While the dense instructions demonstrated consistent 128-cycle latency, the sparse variants exhibit slightly higher latency at 144 clock cycles in ``SS" mode  due to the additional overhead of handling sparsity patterns.




\begin{table*}[!htbp]
    \centering
    \caption{Different sparse \texttt{wgmma} Instructions on H800 Tensor Cores. Latency (LAT) is measured in clock cycles. Throughput is measured in TFLOPS or TOPS/s.}
    \label{tab:sparsewgmma}
    
    \begin{tabular}{lllcccc}
        \toprule
        \textbf{A/B} & \textbf{C/D} & \textbf{Sparse} & \multicolumn{2}{c}{\textbf{LAT/Throughput (Zero)}} & \multicolumn{2}{c}{\textbf{Throughput (Rand)}} \\
        
        \cmidrule(lr){4-5} \cmidrule(lr){6-7}
        
        & & \textbf{Instruction} & \textbf{SS} & \textbf{RS} & \textbf{SS} & \textbf{RS} \\
        \midrule
        
        FP16 & FP16 & \texttt{sp.m64n256k32} & 144.0/1308.0 & 128.0/1472.0 & 1257.8 & 1362.3 \\
        FP16 & FP32 & \texttt{sp.m64n256k32} & 144.0/1312.3 & 128.0/1476.2 & 1194.3 & 1277.5 \\
        TF32 & FP32 & \texttt{sp.m64n256k16} & 144.0/656.8 & 128.0/735.4 & 644.9 & 721.7 \\
        FP8 & FP16 & \texttt{sp.m64n256k64} & 144.0/2619.9 & 128.0/2945.0 & 2588.6 & 2782.4 \\
        FP8 & FP32 & \texttt{sp.m64n256k64} & 144.0/2622.8 & 128.0/2931.0 & 2588.7 & 2722.3 \\
        INT8 & INT32 & \texttt{sp.m64n256k64} & 144.0/2612.4 & 128.0/2933.0 & 2593.9 & 2898.3 \\
        
        \bottomrule
    \end{tabular}
\end{table*}

\noindent\textbf{Matrix Size Scaling Analysis.} Figure 2 illustrates the execution time breakdown for FP8 matrix multiplication using te.Linear across different matrix sizes, demonstrating the performance scaling characteristics of low-precision operations. For the smaller matrix size (N=4096), overhead components dominate the execution profile, with "Other" operations consuming 44.9\% and memory operations (Memcpy) accounting for 29.4\% of total time, while kernel execution represents only 25.3\%. In contrast, the larger matrix size (N=16384) exhibits dramatically improved efficiency, with kernel execution dominating at 84.7\% of total time while overhead components are significantly reduced to 8.2\% for "Other" and 7.1\% for memory operations. This scaling behavior highlights how increased computational intensity at larger problem sizes enables te.Linear to achieve better utilization of tensor cores and amortize the fixed overhead costs, making low-precision formats particularly advantageous for large-scale matrix operations where compute-bound performance can be fully realized.

\begin{figure}[]
    \centering
    \includegraphics[width=0.7\linewidth]{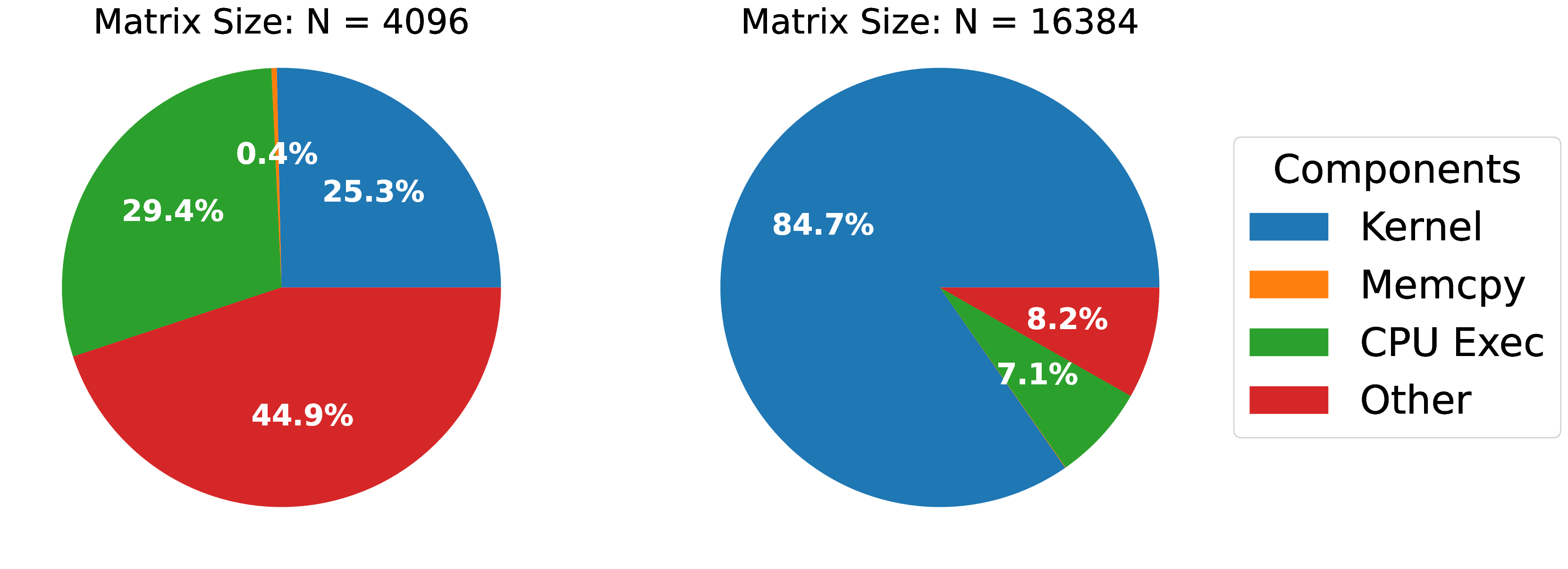} 
    \caption{Proportion of execution time for different operators when performing FP8 matrix multiplication using \texttt{te.Linear}.}
    \label{fig:Linear_pie}
\end{figure}

\end{document}